\documentclass[twocolumn]{aastex61}
\submitjournal{ApJ}

\usepackage{hyperref}
\usepackage{amsmath}
\usepackage{amssymb}
\usepackage{graphicx}
\usepackage{enumitem}
\usepackage{makecell}

\begin{document}
\title{The Lanthanide Fraction Distribution in Metal-poor Stars:\\A Test of Neutron Star Mergers as the Dominant $r$-process Site}
\shorttitle{Lanthanide fractions in metal-poor stars}

\newcommand{\affilcarnegie}{The Observatories of the Carnegie Institution for Science, 813 Santa Barbara St., Pasadena, CA 91101, USA}
\newcommand{\affiljina}{Joint Institute for Nuclear Astrophysics -- Center for Evolution of the Elements, USA}
\newcommand{\affiltoronto}{Department of Astronomy and Astrophysics, University of Toronto, 50 St. George Street, Toronto, Ontario, M5S 3H4 Canada}
\newcommand{\affiltamu}{Mitchell Institute for Fundamental Physics and Astronomy and Department of Physics and Astronomy, Texas A\&M University, College Station, TX 77843-4242, USA}

\newcommand{\Msun}{M_\odot}
\newcommand{\XLa}{X_{\text{La}}}
\newcommand{\XLaAll}{X_{\text{La,\,all}}}
\newcommand{\XLaRed}{X_{\text{La,\,red}}}
\newcommand{\XLaBlue}{X_{\text{La,\,blue}}}
\newcommand{\MAll}{M_{\text{all}}}
\newcommand{\MRed}{M_{\text{red}}}
\newcommand{\MBlue}{M_{\text{blue}}}

\newcommand{\rproc}{$r$-process}

\defcitealias{Arlandini99}{A99}
\defcitealias{Arnould07}{A07}
\defcitealias{Sneden08}{S08}
\defcitealias{Bisterzo14}{B14}
\defcitealias{Roederer18a}{R18}
\defcitealias{Hansen18}{H18}
\defcitealias{Sakari18}{S18}
\defcitealias{jinabase}{JINAbase}

\correspondingauthor{Alexander P. Ji}
\email{aji@carnegiescience.edu}

\author{Alexander P. Ji}
\altaffiliation{Hubble Fellow}
\affiliation{\affilcarnegie}
\affiliation{\affiljina}

\author{Maria R. Drout}
\altaffiliation{Hubble Fellow, Carnegie-Dunlap Fellow, CIFAR Azrieli Global Scholar}
\affiliation{\affiltoronto}
\affiliation{\affilcarnegie}

\author{Terese T. Hansen}
\affiliation{\affiltamu}
\affiliation{\affilcarnegie}

\begin{abstract}
Multimessenger observations of the neutron star merger GW170817 and its kilonova proved that neutron star mergers can synthesize large quantities of $r$-process elements.
If neutron star mergers in fact dominate all $r$-process element production, then the distribution of kilonova ejecta compositions should match the distribution of $r$-process abundance patterns observed in stars.
The lanthanide fraction ($\XLa$) is a measurable quantity in both kilonovae and metal-poor stars, but it has not previously been explicitly calculated for stars. 
Here, we compute the lanthanide fraction distribution of metal-poor stars ([Fe/H]~$< -2.5$) to enable comparison to current and future kilonovae.
The full distribution peaks at $\log \XLa \sim -1.8$, but $r$-process-enhanced stars ([Eu/Fe]~$> 0.7$) have distinctly higher lanthanide fractions; $\log \XLa \gtrsim -1.5$.
We review observations of GW170817 and find general consensus that the total $\log \XLa = -2.2 \pm 0.5$, somewhat lower than the typical metal-poor star and inconsistent with the most highly $r$-enhanced stars.
For neutron star mergers to remain viable as the dominant $r$-process site, future kilonova observations should be preferentially lanthanide-rich (including a population of $\sim 10\%$ with $\log \XLa > -1.5$).
These high-$\XLa$ kilonovae may be fainter and more rapidly evolving than GW170817, posing a challenge for discovery and follow-up observations.
Both optical and (mid-)infrared observations will be required to robustly constrain kilonova lanthanide fractions. 
If such high-$\XLa$ kilonovae are \emph{not} found in the next few years, that likely implies that the stars with the highest $r$-process enhancements have a different origin for their $r$-process elements.
\end{abstract}

\keywords{nuclear reactions, nucleosynthesis, abundances --- stars: neutron --- stars: abundances --- gravitational waves}

\section{Introduction}
Half of the elements heavier than iron are synthesized in the rapid neutron-capture process ($r$-process).
The qualitative nuclear physics of the $r$-process has been understood for decades \citep{Burbidge57, Cameron57}: in extremely neutron-rich environments, the rapid rate of neutron captures exceeds the typical timescale for a nucleus to beta-decay, resulting in three heavy element abundance peaks slightly offset from magic neutron numbers.
However, the astrophysical site (or sites) responsible for producing the majority of these $r$-process elements has been long debated.
The main candidates have been core-collapse supernovae
\citep[e.g.,][]{Woosley92, Surman06, Winteler12, Siegel18}, 
or compact binary mergers involving neutron stars \citep[e.g.,][]{Lattimer74,Lattimer76,Lattimer77}.

We now know for certain that binary neutron star mergers (NSMs) \emph{do} produce $r$-process elements. On Aug 17th, 2017, Advanced LIGO/Virgo observed gravitational waves from the coalescence of two neutron stars (GW170817, \citealt{LIGOGW170817a,LIGOGW170817b}) and a rapidly-evolving optical/infrared transient was subsequently identified \citep{Coulter17}. The qualitative features of this transient closely matched theoretical expectations for a kilonova (KN)---a transient powered by the radioactive decay of newly synthesized $r$-process elements \citep{Drout17, Evans17, Kasliwal17, Smartt17, Cowperthwaite17, Arcavi17, Chornock17,Shappee17,Kilpatrick17,Troja17,Tanvir17,Pian17,Nicholl17,McCully17}. In particular, the observed bolometric light curve, rapid color evolution, and near-infrared spectral features imply the production of high-opacity lanthanides, a smoking gun signal for $r$-process nucleosynthesis \citep[e.g.,][]{Barnes13, Tanaka13, Kasen13}. While large uncertainties remain, the total ejected mass and merger rate inferred from GW170817 imply that NSMs alone may be able to produce enough $r$-process elements to be responsible for all cosmic $r$-process production.

However, it is still unclear if NSMs are the \emph{dominant} source of $r$-process elements.
NSMs have difficulty reproducing chemical evolution trends both in the metal-poor and metal-rich regimes, due to their merging delay times.
At the metal-poor end, it is not clear if NSMs can merge fast enough to produce $r$-process elements at low-enough metallicity \citep[e.g.,][]{Qian00,Argast04,Cescutti15,Wehmeyer15,Safarzadeh19}.
At the metal-rich end, $r$-process element evolution in the Milky Way tends to follow that of core-collapse supernovae \citep[e.g.,][]{McWilliam97, Sneden08}, and adding a delay time distribution to $r$-process production tends to break this match \citep[e.g.,][]{Cote18}.
This issue is by no means settled, as there are a myriad of complicating factors to these first-order arguments.
For example, metal-poor stars can be born in dwarf galaxies with inefficient star formation, mitigating issues with merging delay times \citep{Tsujimoto14b,Ishimaru15,Ji16b,Hansen17,Duggan18} but adding the complication of velocity kicks that should unbind the neutron star binary from the galaxy \citep{Bramante16,Beniamini16,Beniamini18a,Safarzadeh18b,Safarzadeh19}.
Another issue is the role of inhomogenous mixing of $r$-process material, which has a strong impact on observed abundance trends and scatter \citep{Shen15,vandeVoort15,Hirai15,Safarzadeh17a,Naiman18,Schonrich19}.
Several authors have advocated that the dominant $r$-process site transitions from rare classes of supernovae at low metallicity/early times to neutron star mergers at high metallicity/late times \citep[e.g.,][]{Cescutti15, Wehmeyer15, Cote18, Siegel18}.

Despite the uncertainties, it is plausible that neutron star mergers may dominate $r$-process element production over all cosmic time.
If so, the \emph{distribution} of $r$-process compositions observed in metal-poor stars provides baseline expectations for the composition of future KNe. Conversely, KN observations can provide an independent check on the origin of the $r$-process in metal-poor stars.
It is thus important to compare the $r$-process compositions derived from these two different types of observations.
The $r$-process composition of metal-poor stars is thought to trace a small number of (or even individual) $r$-process events, allowing the $r$-process composition to be studied element-by-element \citep[e.g.,][]{Cowan02,Hill02,Sneden03,Sneden08,Roederer12}.
From these studies, it is well-established that stars with $r$-process enhancements display a remarkably robust abundance pattern between the second and third $r$-process peaks \citep{Sneden08}, but there are clear variations in the first peak elements relative to the heavier elements \citep[e.g.,][]{McWilliam95, McWilliam98, Francois07, SiqueiraMello14, Ji18}.
In contrast, only a bulk lanthanide fraction ($\XLa$) can currently be inferred from KNe, although future theoretical development may eventually allow inferences of individual element abundances from infrared spectra or nebular phase emission \citep[e.g.,][]{Kasen17,Smartt17,Kasliwal18}.

In this paper, we compute the lanthanide fraction distribution of metal-poor stars to facilitate comparison between metal-poor stars and KN observations.
We determine the lanthanide fraction distribution from metal-poor stars in Section~\ref{s:starxla}. In Section~\ref{s:kilonova}, we describe how the total lanthanide fraction of NSM ejecta can be measured from observation of kilonovae, focusing on results from the KN following GW170817. Our main results are presented in Section~\ref{s:discussion}, where we compare the lanthanide fraction measured in GW170817 to the distribution observed in metal-poor stars, and consider implications for future kilonova followup and the dominant $r$-process production site. We conclude in Section~\ref{s:conclusion}.

\section{Lanthanide Fraction from Abundance Patterns of Metal-Poor Stars}\label{s:starxla}

The lanthanide fraction ($\XLa$) is the mass ratio between the high-opacity lanthanides (atomic numbers $Z = 57 - 71$) and the total mass.
Neutron star merger ejecta is essentially entirely composed of $r$-process elements. 
Thus, given a full element abundance pattern of a star whose neutron-capture elements can be entirely attributed to one NSM,
it is in principle straightforward to calculate the lanthanide fraction of the associated NSM:
simply take the mass of lanthanides and divide by the mass of all neutron-capture elements.
In practice, this is not possible.
The Sun is the only star with detailed abundances of all $r$-process elements, but the Solar abundances are the product of many Gyrs of chemical enrichment, and a large $s$-process component has to be subtracted off.

The neutron-capture elements of metal-poor halo stars are likely relatively pure tracers of NSM enrichment (e.g., without significant $s$-process contamination for $\mbox{[Fe/H]} < -2.5$, \citealt{Simmerer04}), but in these stars only a relatively small number of neutron-capture elements can be measured.
Our approach is thus to use the neutron-capture element abundances of metal-poor stars, extrapolating the abundance of unmeasured elements using the Solar $r$-process pattern.
Rather than trying to infer all individual elements, we group the neutron-capture elements into three physically motivated mass categories and then take ratios of these mass categories that can be measured from observations (Figure~\ref{f:pattern1}, Table~\ref{tab:categories}).
We subsequently convert these measurements into values of $\XLa$ (Figure~\ref{f:XLa_f}), and derive $\XLa$ distributions from samples of metal-poor stars (Figure~\ref{f:histogram_samples}).

\subsection{Metal-poor star sample selection}\label{s:selection}

\begin{deluxetable*}{l|r|r|r|r}[t]
\tablecaption{Metal-Poor Star Samples \label{tab:samples}}
\tablewidth{0pt}
\tablehead{ \colhead{Sample} & \colhead{Total Number} & \colhead{Metal-poor,} & \colhead{Metal-poor, $r$-dominated,} & \colhead{Metal-poor, $r$-dominated,} \\
            \colhead{} & \colhead{} & \colhead{$r$-dominated} & \colhead{has Sr, Ba, Eu} & \colhead{$\geq$ 5 neutron-capture elements}}
\startdata 
\citetalias{jinabase}    & 426 & 146 & 143$^*$ & 120 \\
\citetalias{Roederer18a}  &  83 &  46 &  45$^*$ &  30 \\
RPA \citetalias{Sakari18} & 125 &  32 & 32$^*$ & 30 \\
RPA \citetalias{Hansen18} & 107 &  39 &  31$^*$ & \ldots \\
\enddata
\tablecomments{Metal-poor: $\mbox{[Fe/H]} < -2.5$; $r$-dominated: $\mbox{[Ba/Eu]} < -0.4$. For the \citetalias{Sakari18} sample, we instead use $\mbox{[Fe/H]} < -2.3$ to account for non-LTE metallicities. The $^*$ symbol indicates which sample is used in figures below.}
\end{deluxetable*}

Our goal of measuring lanthanide fractions requires a large, complete, and pure sample of metal-poor stars.
By ``complete,'' we mean that our sample should be unbiased with respect to $r$-process-enhancement.
The $r$-process-enhancement level of a star is typically quantified by [Eu/Fe], as Eu has strong stellar absorption features and is a relatively clean tracer of the $r$-process.
By ``pure,'' we mean that the stars have neutron-capture elements dominated by the $r$-process.
The primary contaminant, the $s$-process, can be mostly removed by a cut on $\mbox{[Ba/Eu]} < 0$ \citep{Beers05}\footnote{Eu is created by both the $r$- and $s$-processes, but in the $s$-process it is also accompanied by a large amount of Sr and Ba (representing the first and second $r$-process peaks). The [Ba/Eu] ratio thus roughly quantifies the extent of $s$-process contamination in a star. This separation could also be done with Pb \citep{Roederer10b}, but few stars have Pb constraints.}, as well as $\mbox{[Fe/H]} < -2.5$ \citep{Simmerer04}.
We thus consider ``metal-poor stars'' to be stars with $\mbox{[Fe/H]} < -2.5$, which is also the regime where it's plausible (though not guaranteed) that most or all of the $r$-process content of a star comes from a single $r$-process enrichment event.

We will primarily consider two different data samples that trade off purity and completeness.
First, we use the $r$-process star sample from \citet{Roederer18a} (\citetalias{Roederer18a}).
This sample compiles 83 field stars with $\mbox{[Eu/Fe]} > 0.7$, which we will call $r$-enhanced stars\footnote{Traditionally, $r$-enhanced stars have been divided into two sub-categories of $r$-I stars ($0.3 < \mbox{[Eu/Fe]} < 1.0$) and $r$-II stars ($\mbox{[Eu/Fe]} > 1.0$) \citep{Beers05}, but for clarity we will use the single separation threshold defined by \citetalias{Roederer18a} in this paper}.
These 83 stars have received individual attention, guaranteeing that their heavy elements are dominated by the $r$-process (and thus purity); but they are purposely biased to high [Eu/Fe] compared to typical halo stars.
We collect the neutron-capture element abundances of these stars from the literature.
Second, we use the \citetalias{jinabase} database of metal-poor star abundances \citep{jinabase}.
This is a large inhomogeneous literature compilation with an ambiguous selection function, but it is intended to be complete for literature stars with $\mbox{[Fe/H]} < -2.5$ prior to 2015.
It contains most (but not all) of the stars in the \citetalias{Roederer18a} sample.
We only consider halo stars from \citetalias{jinabase} (i.e., removing dwarf galaxy and bulge stars) that have Eu measurements, resulting in a total sample of 426 stars.
(See Appendix~\ref{litrefs} for associated literature citations.)

To obtain a metal-poor and pure $r$-process star sample, we apply cuts of $\mbox{[Fe/H]} < -2.5$ and $\mbox{[Ba/Eu]} < -0.4$.
The [Ba/Eu] cut is chosen to be more stringent than usual, because we found the less stringent cut did not remove all stars with $s$-process signatures in the JINAbase sample.
This reduces the sample sizes to 146 in JINAbase and 46 in R18, of which nearly all stars have Sr, Ba, and Eu measured (143 and 45 stars, respectively).
If we further restrict to stars with ${\geq}5$ neutron-capture elements, the final sample sizes are 120 stars in JINAbase and 30 stars from R18.

As a separate check, we also consider two complete and uniformly analyzed samples from the $R$-Process Alliance (RPA, \citealt{Hansen18,Sakari18}, abbreviated \citetalias{Hansen18,Sakari18}).
These two samples are much smaller than (and have partial overlap with) JINAbase, but were designed to have an unbiased selection in [Eu/Fe]\footnote{Specifically targeting bright, cool, metal-poor stars, guaranteeing [Eu/Fe] measurements or useful limits}.
After we apply [Fe/H] and [Ba/Eu] cuts, each RPA sample contains ${\sim}30$ stars with Sr, Ba, and Eu.
For the \citetalias{Sakari18} sample, we use a different $\mbox{[Fe/H]} < -2.3$ cut, to account for the fact that those metallicities were determined with non-LTE corrections.
We will find that the $\XLa$ results from the RPA samples match those from the JINAbase sample, suggesting that the inhomogeneous literature data in JINAbase turns out to be a fair selection.
Note that for all samples, we ignore stars with Eu upper limits, because we cannot be certain that they are $r$-process dominated. 
This may result in some bias to the inferred $\XLa$ distributions, though the number of limits is small for the RPA samples.
An overview of the samples and selection criteria is given in Table~\ref{tab:samples}.

\subsection{Physically motivated split of $r$-process elements}\label{s:categories}
\begin{table*}
    \centering
    \begin{tabular}{c|ccc}
         \hline
         Mass category & Physical Conditions & \multicolumn{2}{c}{Description}  \\
         \hline
         $M_A$ & $Y_e \gtrsim 0.25$, $\kappa \lesssim 1$  & \multicolumn{2}{c}{first $r$-process peak, $70 \leq A < 115$} \\
         $M_B$ & $Y_e \lesssim 0.25$, $\kappa \lesssim 1$  & \multicolumn{2}{c}{second and third $r$-process peaks, $115 \leq A < 140$ and $176 \leq A < 211$}  \\
         $M_C$ & $Y_e \lesssim 0.25$, $\kappa \gtrsim 10$ & \multicolumn{2}{c}{lanthanides and actinides, $140 \leq A < 176$ and $A > 230$} \\
         \hline
         Ratio & Equation & Value & Description \\
         \hline
         $H$    & $\frac{M_C}{M_B+M_C}$      & $0.14 \pm 0.02$   & Measured in Sun, set by nuclear physics \\
         $f$    & $\frac{M_A}{M_B+M_C}$      & Varies (Section~\ref{s:fdefined}) & Measured in Sun and metal-poor stars, set by astrophysics \\
         $\XLa$ & $\frac{M_C}{M_A+M_B+M_C}$  & $10^{-2.2 \pm 0.5}$ & Measured in kilonovae, set by nuclear physics and astrophysics \\
    \end{tabular}
    \caption{Description of different mass categories (see Fig~\ref{f:pattern1}) and mass ratios (see Sections~\ref{s:categories}, \ref{s:observedvalues}). The exact values of $Y_e$ and $\kappa$ for physical conditions should not be taken literally.}
    \label{tab:categories}
\end{table*}

The details of $r$-process nucleosynthesis and radiative transfer in kilonovae are complex, but two bulk patterns are well-established.
First, the result of $r$-process nucleosynthesis in a mass component is primarily determined by its electron fraction $Y_e \equiv n_e/(n_n + n_p)$ \citep[e.g.,][]{Lippuner15}.
Neutron-poor, or high $Y_e$ ejecta ($Y_e \gtrsim 0.25$) primarily form $r$-process elements in the first peak; while neutron-rich, or low $Y_e$ ejecta ($Y_e \lesssim 0.25$) primarily form $r$-process elements in the second peak and above, including all of the lanthanides.
Second, the opacity of the lanthanides and actinides is ${\gtrsim}10{\times}$ higher than other elements \citep[e.g.,][]{Kasen13, Fontes15,Tanaka13}. 
These elements have open $f$-shells (i.e. are in the two split-off rows in the periodic table), so they have many electronic transitions and thus a high line density at optical wavelengths. 
As a result, if synthesized, lanthanides will dominate the opacity in kilonovae--making their presence in the ejected material possible to infer (see Section~\ref{sec:KNtheory}).

\begin{figure}
\centering
\includegraphics[width=1\linewidth]{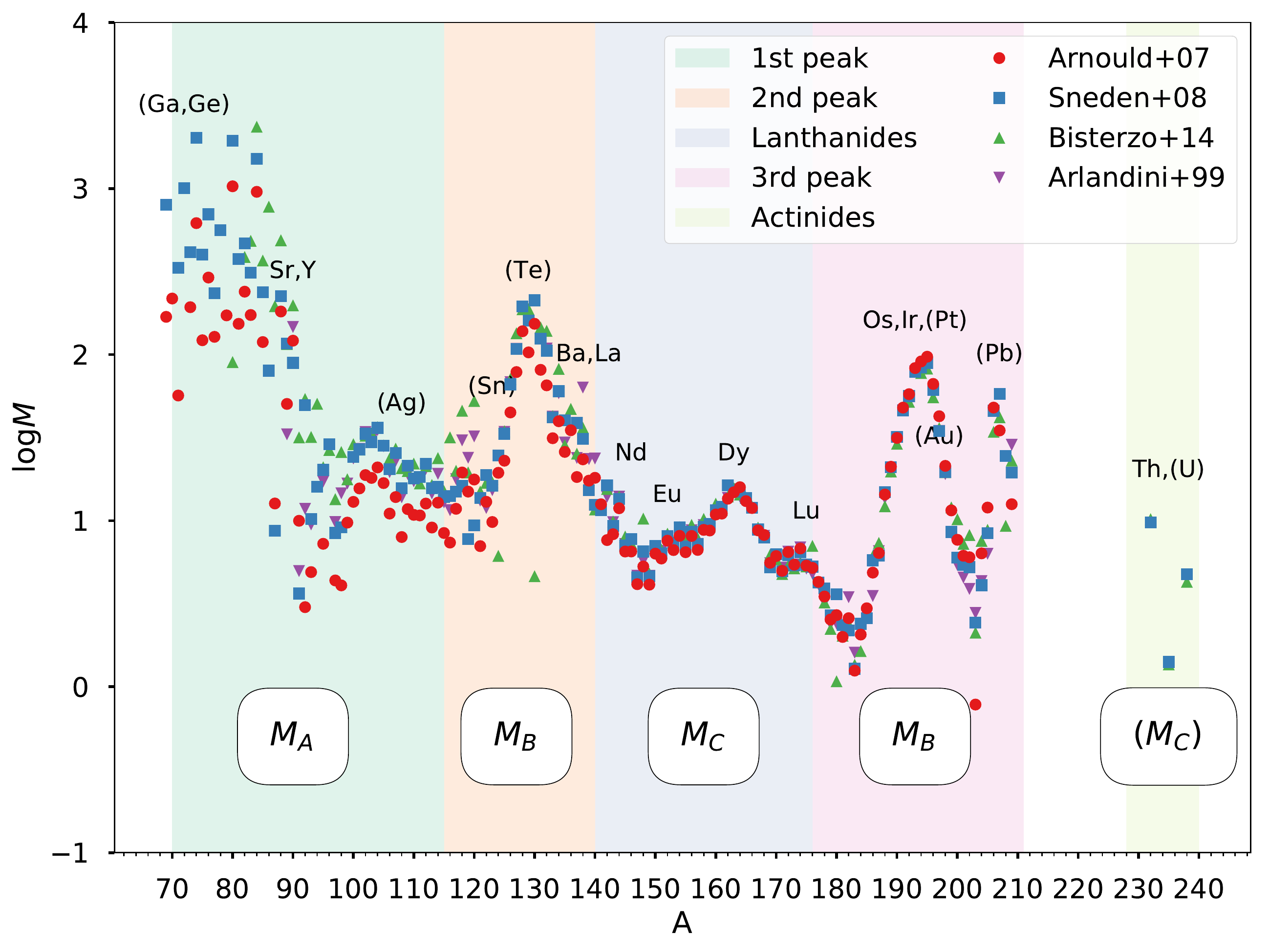}
\caption{
$\log M$ vs atomic mass $A$ for the solar $r$-process pattern, according to four different models of $s$-process subtraction (colored dots, \citetalias{Arnould07,Sneden08,Bisterzo14,Arlandini99}).
The overall scaling of $\log M$ is arbitrary.
The shaded regions indicate how we physically divide the isotopes into groups, and labeled squares at the bottom show how we separate the isotopes into categories $M_A$, $M_B$, and $M_C$. Gaps at $210 < A < 228$ and $A > 240$ indicate isotopes that decay within 1 Gyr. We will not include the actinides in any of our calculations.
The solar $r$-process pattern is very consistent starting from the 2nd peak, but there is significant variation in the first peak due to differences in how the $s$-process is subtracted.
Key elements measured in metal-poor stars are labeled here. Elements in parentheses are measured in $\lesssim 5$ metal-poor stars, because their lines are weak/highly blended, and/or require UV observations. We will use the \citetalias{Arnould07,Sneden08} solar patterns to determine the abundance of missing elements, as the other two patterns do not reach isotopes with $A < 85$.
\label{f:pattern1}}
\end{figure}

Based on these considerations, we divide the $r$-process composition into three categories denoted with masses $M_A$, $M_B$, and $M_C$ (see Table~\ref{tab:categories}, Figure~\ref{f:pattern1}).
Category $M_A$ is composed of first peak elements, which are synthesized in high-$Y_e$ material and have low opacity.
Category $M_B$ is composed of second and third peak elements, which are synthesized in low-$Y_e$ material but have low opacity.
Category $M_C$ is composed of the lanthanides and actinides, which are synthesized in low-$Y_e$ material and have high opacity.
In this paper, we neglect the actinides, as they are poorly constrained by abundance observations in stars and the abundance is low enough that it should not substantially affect our conclusions (though see Section~\ref{s:compareKNstar}).

We then define three mass ratios that are measured independently:
\begin{align}
    H &\equiv \frac{M_C}{M_B+M_C} \\
    f &\equiv \frac{M_A}{M_B+M_C} \\
    \XLa &\equiv \frac{M_C}{M_A+M_B+M_C}
\end{align}
$H$ relates the high-opacity lanthanides to the rest of the elements synthesized in neutron-rich ejecta. It traces the so-called ``universal $r$-process,'' which has shown to be essentially identical in the solar $r$-process residuals and in metal-poor stars \citep[e.g.,][]{Sneden08}.
$f$ is the ratio between neutron-poor ejecta and neutron-rich ejecta, which traces astrophysics of the neutron star merger and can be measured in metal-poor stars \citep[e.g.,][]{Ji18}.
As described above, $\XLa$ is the ratio between the high-opacity lanthanides and the total mass of $r$-process material, and it heavily influences the observed color and duration of a kilonova \citep[e.g.,][]{Kasen17}.
These three ratios are not independent but are related by this equation:
\begin{equation}\label{eq:ratioequation}
    \XLa = \frac{H}{1+f}
\end{equation}

For simplicity, in this paper we have separated the mass categories $M_A$, $M_B$, and $M_C$ with hard cuts based on atomic number (Table~\ref{tab:categories}, Figure~\ref{f:pattern1}).
This is an imperfect simplification, especially for our cutoff of $A=115$ between the first and second $r$-process peaks. Full $r$-process nucleosynthesis calculations show a fuzzier boundary that depends on the exact value of $Y_e$, as well as the ejecta entropy and expansion timescale \citep[e.g.,][]{Lippuner15}.
In particular, high $Y_e$ ejecta can produce elements in (but not beyond) the second $r$-process peak, effectively transferring mass from $M_A$ to $M_B$.
However, the parameter range for this overlap is relatively small, so we do not expect this choice to affect our results, as long as the exact threshold of $Y_e = 0.25$ is not taken too seriously.
We test that applying ${\pm}5$ amu variations to the $A=115$ cutoff does not affect our results.
Fitting full $r$-process nucleosynthesis models to star abundance patterns \citep[e.g.,][]{Hill17} is beyond the scope of this paper. But we note such modeling is unwarranted for most stars (which have too few elements measured for such a fit to be meaningful), and furthermore the predictions have significant theoretical uncertainties \citep[e.g.,][]{Eichler15,MendozaTemis15,Barnes16,Mumpower16,Nishimura16,Vassh18}.
Until better constraints are obtained for the nuclear data, we believe our empirical $\XLa$ calculation based on the Solar pattern is comparably accurate to detailed abundance modeling.

\subsection{Measured values of mass ratios}\label{s:observedvalues}
We now discuss the observed values of $H$ and $f$ in stars. This can be used to infer $\XLa$ in metal-poor stars based on Equation~\ref{eq:ratioequation}.

\subsubsection{$H$: universal $r$-process ratio}
The $r$-process pattern from Ba to Ir ($Z=56-77$, $A \sim 135-195$) has been found to be universal, i.e. the elements have identical abundance ratios in the Sun and in metal-poor halo stars to within measurement uncertainties \citep{Sneden08}.
This universality is thought to be set primarily by nuclear physics processes like fission cycling or nuclear mass structure, though the exact mechanism is not completely understood \citep[e.g.,][]{Eichler15,MendozaTemis15,Nishimura16,Mumpower16}.
The observed universality may not extend \emph{into} the second peak (to elements like Te, I, Xe):
observationally, there are very few direct measurements of elements in this peak for metal-poor stars, and theoretically it is known that different fission fragment models impact the ratio of the second peak to lanthanides \citep[e.g.,][]{Eichler15,Vassh18}.
Since the physics of fission fragments is uncertain and we cannot currently constrain such variations directly in stars, we will assume universality extends into the second peak.
This means we assume the ratio $H = M_C/(M_B + M_C)$ is the same in all stars, and can be precisely measured from the solar system $r$-process isotopic abundances.

The solar system $r$-process pattern is not directly observed, but instead is derived by subtracting off a modeled $s$-process contribution from the actually observed isotopic abundances \citep[e.g.,][]{Arlandini99}.
On the whole this is quite reliable because nuclear data for the $s$-process is well-measured from laboratory experiments, but the $s$-process yield is metallicity dependent due to the varying number of neutron seeds \citep[e.g.,][]{Fishlock14}.
There is thus some variation in the resulting $r$-process pattern depending on the assumed chemical evolution.
In Figure~\ref{f:pattern1}, we show four different $r$-process patterns from \citet{Bisterzo14,Arlandini99,Arnould07,Sneden08} \citepalias[abbreviated ][]{Bisterzo14,Arlandini99,Arnould07,Sneden08}.
For \citetalias{Bisterzo14,Arlandini99}, we apply their isotopic fractions to the solar abundances by \citet{Lodders10}.
These four different patterns give $H$ ranging from $0.13-0.15$ with a statistical error of $0.01$, using the atomic mass ranges in Table~\ref{tab:categories}.
Shifting our atomic mass ranges around for $M_B$ results in only $0.01$ additional uncertainty on $H$.
We thus adopt $H=0.14 \pm 0.02$ for all stars.

\subsubsection{$f$: neutron-richness ratio}\label{s:fdefined}
In principle, $f$ can also be calculated from the solar $r$-process pattern the same way as $H$.
However, there are significant differences between the models in the first peak of the solar $r$-process pattern (Figure~\ref{f:pattern1}).
The reason is that the first peak elements are more affected by uncertainties in the $s$-process subtraction compared to the heavier elements/isotopes \citep[see e.g.,][]{Sneden08,Bisterzo14}.
Furthermore, unlike $H$, the abundance of the first peak elements varies significantly in metal-poor stars compared to the heavier $r$-process elements \citep[e.g.,][]{McWilliam98,Sneden08,SiqueiraMello14,Ji18}.

We thus calculate $f$ from a combination of the solar abundance pattern and element abundances in metal-poor stars.
Specifically, we calculate $M_{A,\odot},\,M_{B,\odot},\,M_{C,\odot}$ from the \citetalias{Arnould07} and \citetalias{Sneden08} solar $r$-process patterns (the other two patterns do not extend to the beginning of the first peak).
The overall normalization does not matter, as only the ratio $M_{A,\odot}/(M_{B,\odot}+M_{C,\odot})$ will affect $f$.
This ratio is 2.45 and 4.66 for \citetalias{Arnould07} and \citetalias{Sneden08}, respectively, which is dominated by differences in the first peak elements (Figure~\ref{f:pattern1}).
Then, for each star in the samples of $r$-process enhanced metal-poor stars described above, we calculate $\Delta\log\epsilon(A)$ and $\Delta\log\epsilon(BC)$, the offsets required to minimize the absolute deviation between the observed abundances and the solar number abundance, i.e.
\begin{equation}\label{eq:deltalogeps}
    \Delta\log\epsilon(A) = \text{argmin}_\Delta \sum_Z \left|\log\epsilon(Z) - \log\epsilon_{r,\odot}(Z) + \Delta\right|
\end{equation}
For $\Delta\log\epsilon(A)$ we take elements $31 \leq Z \leq 47$, and for $\Delta\log\epsilon(BC)$ we take $56 \leq Z \leq 77$.
Finally, we obtain
\begin{equation}\label{eq:fcalc}
    f = \frac{M_{A,\odot}}{M_{B,\odot}+M_{C,\odot}}10^{\Delta\log\epsilon(A) - \Delta\log\epsilon(BC)}
\end{equation}
This procedure essentially calculates $f$ of individual metal-poor stars, using the solar $r$-process pattern to extrapolate element abundances that are not observed in those stars, explicitly allowing for separate normalizations of the light (A) and heavy (BC) $r$-process elements.
The typical scatter in the offsets is 0.2 dex (as measured by the biweight standard deviation), which we adopt as the typical uncertainty for $\log f$.
We consider this procedure most reliable when ${\geq}5$ neutron-capture elements are measured (i.e. at least 2 elements for $A$ and 3 elements for $B$+$C$), but this biases us to stars with higher [Eu/Fe]. Thus, we will calculate $\XLa$ stars with ${\geq}3$ elements (i.e. one element for $A$, typically Sr, and two elements for $B$+$C$, typically Ba and Eu).

Note that our $\XLa$ calculation weights all element abundances equally. 
In principle, some elements are better-measured than other, so one should weight by the abundance uncertainty in Equation~\ref{eq:deltalogeps} (e.g. as $1/\sigma_Z^2$ for a standard $\chi^2$, or $1/\sigma_Z$ for a weighted absolute deviation).
In practice, literature compilations have different abundance error definitions that preclude uniform application of such weights.
However, we find that using the weighted absolute deviation does not significantly affect our final $\XLa$ distributions for either the \citetalias{Roederer18a} sample or a subset of the JINAbase sample with $\mbox{[Eu/Fe]} > 0.3$.

\begin{figure}[b]
\centering
\includegraphics[width=.95\linewidth]{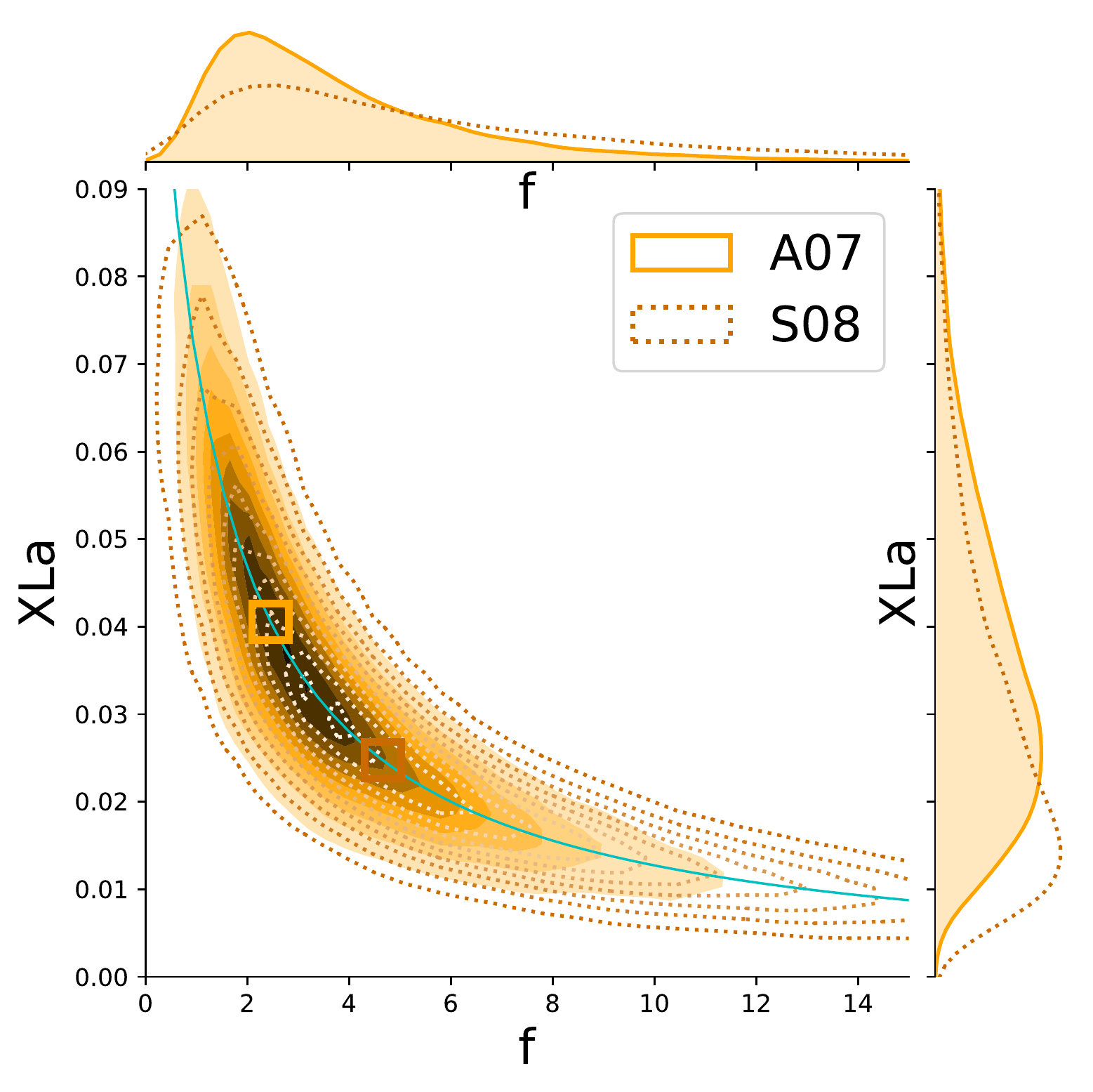}
\caption{
Distribution of $\XLa$ using $H$ and $f$ from $r$-process measurements in stars for the R18 sample.
Center panel shows the joint distribution, with 1D marginal distributions on the other two axes.
We have used $H = 0.14 \pm 0.02$, and 0.2 dex error for $\log f$.
Orange filled contours and brown dotted contours indicate results using the solar patterns from \citetalias{Arnould07} and \citetalias{Sneden08}, respectively.
The open orange and brown squares indicate $f$ and $\XLa$ from the solar composition.
The solid cyan line shows $\XLa = H/(1+f)$ line for $H=0.14$.
\label{f:XLa_f}}
\end{figure}

\begin{figure*}[t]
\centering
\includegraphics[width=\linewidth]{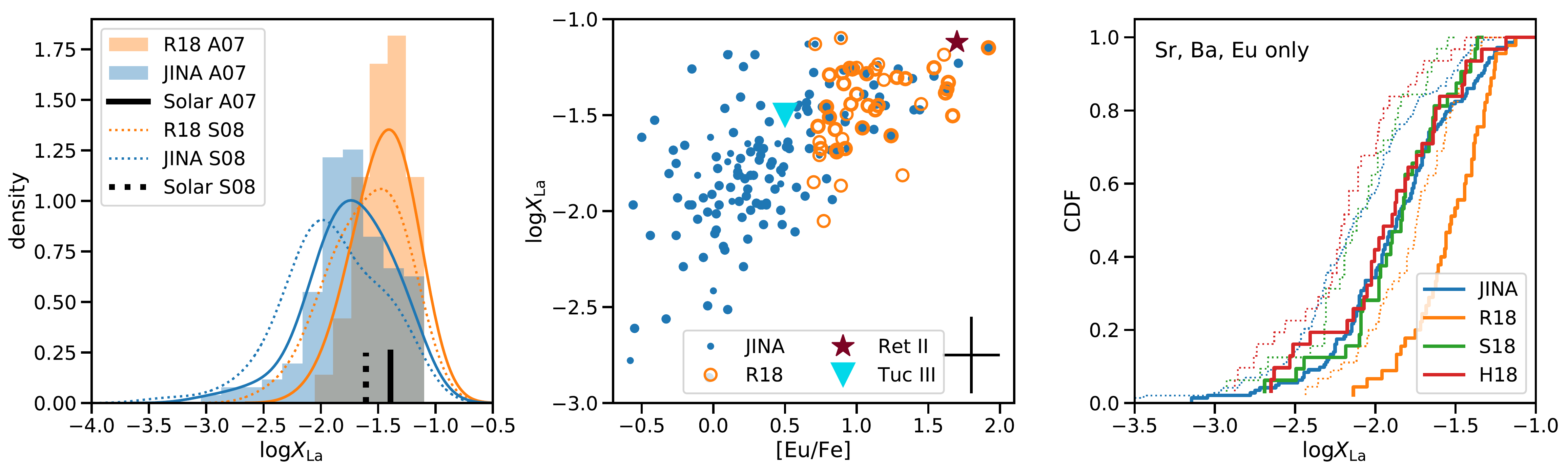}
\caption{
\emph{Left:}
$\log\XLa$ distributions for \citetalias{jinabase} (blue) and \citetalias{Roederer18a} (orange) samples of stars with ${\geq}3$ neutron-capture elements measured.
The shaded histograms are for the \citetalias{Arnould07} solar pattern, with the corresponding kernel density estimate in solid lines (width=0.2 dex).
The dotted lines show $\XLa$ for the same samples but computed with the \citetalias{Sneden08} solar pattern.
The $r$-enhanced R18 sample has distinctly higher $\XLa$.
For comparison, the Solar lanthanide fraction is shown as a short black solid (dotted) line for the \citetalias{Arnould07} (\citetalias{Sneden08}) solar pattern.
\emph{Middle:} $\XLa$ vs [Eu/Fe] for stars with ${\geq}3$ neutron-capture elements measured (same samples as left panel).
Stars with ${\geq}5$ neutron-capture elements measured are in bolder symbols.
The black cross indicates approximate error bars. $\XLa$ is computed with the \citetalias{Arnould07} solar pattern. There is a clear correlation between $\XLa$ and [Eu/Fe].
For comparison, we also show the lanthanide fractions inferred for the $r$-enriched ultra-faint dwarf galaxies Reticulum~II and Tucana~III as a red star and cyan triangle, respectively.
\emph{Right:} cumulative distribution of $\log\XLa$ for four samples of \citetalias{jinabase}, \citetalias{Roederer18a}, and the two RPA samples from \citet{Sakari18} (S18) and \citet{Hansen18} (H18).
The solid lines are for \citetalias{Arnould07} solar abundances, and the dotted lines are for \citetalias{Sneden08} solar abundances.
We use only Sr, Ba, and Eu to calculate $\XLa$ here, so every star is evaluated with the same elements.
The JINAbase and RPA samples have very similar $\XLa$ distributions.
\label{f:histogram_samples}}
\end{figure*}

A systematic effect can arise if the solar $r$-process pattern is not a good match to the observed abundance pattern in metal-poor stars.
This does not significantly affect the heavier elements in $M_B$ and $M_C$ due to the universal $r$-process pattern, but it does affect the first peak and $M_A$.
In particular, the element yttrium (Y) is often measured in metal-poor stars, but is a poor overall match to stellar abundance patterns for both the \citetalias{Arnould07} and \citetalias{Sneden08} solar patterns that we use here.
This difference in Y goes away with the \citetalias{Bisterzo14} pattern, suggesting the discrepancy is due to how the $s$-process is subtracted.
Our 0.2 dex uncertainty on $\log f$ accounts for such systematic deviations.

Otherwise, we expect this procedure for calculating $f$ to overestimate $M_A$ and $f$ (and thus underestimate $\XLa$) for two reasons.
First, the solar $r$-process residuals used here may have too much mass in $M_{A,\odot}$.
Only three stars in our samples have measurements in the first $r$-process peak (Ge and Ga, $70 < A < 80$), but those measurements are ${\sim}$1 dex below the solar $r$-process pattern.
Theoretical nucleosynthesis calculations support that the $r$-process falls off rapidly below $A \sim 80$ compared to the solar patterns used here \citep[e.g.,][]{Lippuner17}. The \citetalias{Arnould07} $r$-process pattern fits these $Z=31-32$ elements much better, so we prefer this pattern over the \citetalias{Sneden08} pattern.
Second, there are likely sources of neutron-capture element enrichment that produce first peak elements but not the heavier $r$-process elements, such as a neutrino-driven wind in core-collapse supernovae \citep[e.g.,][]{Travaglio04, Montes07, Honda07, Arcones07, Arcones11, Wanajo13}.
This may cause the solar $M_A$ to be too high (as such an additional source has not been subtracted from the observed solar abundances; e.g., \citealt{Travaglio04,Montes07}), and such sources could also possibly contaminate the first peak elements observed in metal-poor stars (see Section~\ref{s:discussionsystematics}).

\subsection{$\XLa$ distribution from metal-poor stars}\label{s:xladistr}

In Figure~\ref{f:XLa_f}, we illustrate the inferred $\XLa$ from the R18 sample using Equation~\ref{eq:ratioequation} and two different solar patterns \citetalias{Arnould07} and \citetalias{Sneden08}.
We use Gaussian uncertainties around $H = 0.14 \pm 0.02$ and 0.2 dex in $\log f$, and Monte Carlo 1000 points for each star to generate these histograms.
Overall the \citetalias{Sneden08} solar pattern is biased to higher $f$ (and thus lower $\XLa$) because the extrapolated mass of first peak elements is higher by a factor of ${\approx} 2$.

Our main $\XLa$ distribution results are shown in Figure~\ref{f:histogram_samples}.
The left panel shows the inferred $\log\XLa$ distributions for the R18 (orange), and JINAbase (blue) samples, for stars with ${\geq}3$ measured neutron-capture elements. The shaded histograms indicate our calculation with the \citetalias{Arnould07} solar pattern, and the solid lines indicate a Gaussian kernel density estimate with width 0.2 dex (the typical error in $\log\XLa$). The dotted lines indicate the kernel density estimates with the \citetalias{Sneden08} pattern.
The full range of values observed in the JINAbase sample is $-3.0 < \log\XLa < -1.1$, peaking at $\log\XLa = -1.71 (-1.95)$ for the \citetalias{Arnould07} (\citetalias{Sneden08}) solar pattern, with a typical scatter of 0.5 dex.
Thus, the majority of stars have lanthanide fractions $\log\XLa \approx -2 \pm 0.5$.
However, the $r$-enhanced stars from the R18 sample, which are canonically used as paragons of $r$-process element enrichment, have significantly higher lanthanide fractions, peaking at $\log\XLa = -1.44 (-1.55)$ for the \citetalias{Arnould07} (\citetalias{Sneden08}) solar pattern, with a typical scatter of 0.3 dex. (The two-sample Kolmogorov-Smirnov test gives $p \lesssim 10^{-5}$ that the $\XLa$ distributions are the same.)
The middle panel of Figure~\ref{f:histogram_samples} shows that the difference between these samples can be explained by a strong correlation between $\log\XLa$ and [Eu/Fe].
This is a new view on an old result: it is well-established that the stars with highest $r$-enhancements are moderately deficient in first peak elements, and thus they must have higher lanthanide fractions for their total $r$-process content (or equivalently, have low [Sr/Ba] and [Sr/Eu] ratios; e.g., \citealt{Travaglio04,Francois07,SiqueiraMello14,Ji18}).

To illustrate that the JINAbase sample is not affected by selection effects,
the right panel of Figure~\ref{f:histogram_samples} shows the cumulative distribution of $\log\XLa$ computed just from Sr, Ba, and Eu, for the \citetalias{jinabase}, \citetalias{Roederer18a}, and RPA (\citetalias{Sakari18,Hansen18}) samples. Solid/dotted lines show results with the \citetalias{Arnould07}/\citetalias{Sneden08} solar abundances.
There is a good match between the RPA and JINAbase samples, and the CDFs also more clearly indicate the stark difference between the $\XLa$ distribution of all metal-poor stars and the $r$-enhanced \citetalias{Roederer18a} stars.
There are some discrepancies in the distribution tails, most notably that JINAbase over-represents the most Eu-enhanced (and thus most $\XLa$-enhanced) stars. This is likely due to publication bias: the most $r$-enhanced stars tend to be singled out for followup and analysis, while larger but more-complete samples of stellar abundances take longer to publish. We thus use the RPA sample distributions when considering the distribution tails.

Another useful comparison is to the $\XLa$ of stars in $r$-process-enriched ultra-faint dwarf galaxies.
The stars in these galaxies preserve a pure signature of an individual $r$-process enrichment event \citep{Ji16b}.
So far, two such galaxies have been found: the highly $r$-enhanced galaxy Reticulum~II \citep{Ji16b,Roederer16b} and the moderately $r$-enhanced galaxy Tucana~III \citep{Hansen17,Marshall18}. We computed $\XLa$ for 7 stars in Reticulum~II \citep{Ji16c} and 4 stars in Tucana~III \citep{Marshall18} the same way as the halo star samples, using the \citetalias{Arnould07} solar pattern. We find $\log\XLa=-1.12 \pm 0.13$ for Reticulum~II and $\log\XLa=-1.50 \pm 0.10$ for Tucana~III. These are shown in the center panel of Figure~\ref{f:histogram_samples}, and they tend to have higher $\XLa$ than halo stars with comparable [Eu/Fe].
It is unclear if the higher $\XLa$ is due to most halo stars being contaminated by additional sources of neutron-capture elements, or intrinsic variation within a single $r$-process site \citep[][also see discussion in Section~\ref{s:discussionsystematics}]{Ji16c,Ji18}.

\subsection{Other potential systematics with metal-poor star lanthanide fractions}\label{s:discussionsystematics}

We re-emphasize that the elements dominating the mass in the 1st and 2nd peaks are usually not directly measured in metal-poor stars (Figure~\ref{f:pattern1}).
We have extrapolated the abundances of these unmeasured elements using the Solar $r$-process pattern (Section~\ref{s:observedvalues}), but this is quite uncertain.
Recall that the Solar 1st $r$-process peak mass is likely over-estimated;
we have assumed universality between the mass ratios of the lanthanides and the 2nd/3rd peak (i.e., $H$ is constant); 
and we have used sharp atomic mass cutoffs that do not capture a fuzzy boundary between the 1st and 2nd $r$-process peaks.
More detailed model fitting to exact stellar abundances could help remedy some of these concerns \citep[e.g.,][]{Hill17, Holmbeck19}, but this is likely only to be possible for stars with large $r$-process enhancements (high [Eu/Fe] values), which is a biased probe of all $r$-process element production (Figure~\ref{f:histogram_samples}, center panel).
Ultimately, removing such assumptions will require directly measure the abundances of more elements in the 1st and 2nd $r$-process peaks with a large sample of stars.
Unfortunately, those elements (e.g., Ga, Ge, Te) require high-resolution ultraviolet spectroscopy \citep{Roederer12}, and significantly expanding beyond the existing sample likely requires future large space-based UV spectrographs like LUMOS/LUVOIR \citep{luvoir}.

Our sample selection criteria effectively removes stars contaminated by the $s$-process, but this may not be the only contaminating source of neutron-capture elements at low metallicity.
Most importantly, there is strong evidence for a source producing only first $r$-process peak that operates in the very early universe \citep{Frebel05,Travaglio04,Montes07,Honda07,JHansen14}.
Stars displaying this abundance signature are called ``weak-$r$'' or ``limited-$r$'' sources \citep{Frebel18ARNPS}, although it is not clear if this pattern is in fact associated with a partial $r$-process or just a neutron-rich freeze out.
If such sources significantly contaminate our sample of metal-poor stars, they would increase our $M_A$ without affecting $M_B+M_C$, thus causing our $\XLa$ inferences to be too low.
There may also be similar contributions from other non-standard neutron-capture element processes at low metallicity \citep[e.g.,][]{Clarkson18,Banerjee18a}.
The fact that the pure $r$-process lanthanide fractions in ultra-faint dwarf galaxies are somewhat higher than bulk halo stars (Figure~\ref{f:histogram_samples}, center panel) may suggest that contamination exists in our halo star sample.

One must also be concerned that metal-poor stars probe a biased sample of $r$-process events.
For example, if NSM lanthanide fractions are covariant with the merger delay time or merger explosion energy, or binary NS system properties vary with metallicity, then metal-poor star lanthanide fractions (which sample star formation at very early times $z \gg 1$) could differ from the overall KN lanthanide fractions (which represent star formation at $z \sim 0$). We do not expect these to be important effects, but this certainly deserves attention from detailed models.

Finally, we have completely neglected actinides in calculating our lanthanide fractions, but they contribute to opacity in a KN. We discuss this more in Section~\ref{s:compareKNstar}, but including actinides would strictly increase $\log\XLa$ inferred for all metal-poor stars, because it adds high-opacity material that has decayed away by the time we observe the stars.

\section{Lanthanide Fractions from Kilonovae}
\label{s:kilonova}

The lanthanide fraction distribution of metal poor stars can test the dominant $r$-process site through \emph{comparison} to independent measures of lanthanide fractions. \emph{Kilonovae} (KNe) are transients powered by the radioactive decay of unstable $r$-process nuclei. When combined with gravitational waves, KNe can provide a distribution of lanthanide fractions where the origin of the $r$-process elements is unambiguously neutron star mergers.

Detailed observational signatures of KN are dependent on the physical parameters of the material, such as the mass, velocity, composition, and geometry. In the context of NSMs, multiple sources of ejected material with different physical parameters are possible: from dynamically ejected tidal tails and material shocked and ejected from the merger interface to accretion disk winds. The physical parameters of each of these theoretical components are predicted to vary both with the binary properties (mass ratio, total mass, eccentricity) and with the nature and lifetime of the merger remnant. Given the number of free parameters, kilonovae from NSMs are expected to be diverse. In addition, even once the physical properties of an ejecta component have been established, accurately translating these properties into observed light curves and spectra requires knowledge of the heating rate from the radioactive decay of $r$-process elements, thermalization efficiency over time, and the opacity of the material. 
Here, we review kilonova observations of GW170817, focusing on ejecta \emph{composition} constraints that can be made via observations, with the goal of directly comparing to the distribution of lanthanide fractions in metal-poor stars derived above.
We summarize salient aspects of KN modeling and $\XLa$ measurements (Section~\ref{sec:KNtheory}), compile and compare literature measurements of GW170817 (Section~\ref{sec:GW170817data} and Figure~\ref{fig:gw170817}), and briefly discuss $\XLa$ measurements in other short gamma ray bursts (Section~\ref{s:sGRB}).
Readers familiar with KNe and GW170817 can skip to Section~\ref{s:compareKNstar}, where we compare KNe to metal-poor stars.

\subsection{Measuring lanthanide fractions from observations of kilonovae} \label{sec:KNtheory}

\begin{deluxetable*}{c|ccc}[t]
\tablecaption{Kilonova Lanthanide Fraction Models \label{tab:kilonovamodels}}
\tablewidth{0pt}
\tablehead{ \colhead{Model} & \colhead{\citet{Kasen17}} & \colhead{\citet{Tanaka18}} & \citet{Wollaeger18}}
\startdata 
Radiative Transfer & \makecell{{\sc{sedona}} \\\citep{Kasen06}} & \makecell{custom 3D Monte-Carlo \\ \citep{Tanaka13}} & \makecell{SuperNu\\\citep{Wollaeger14}} \\
Atomic Data & \makecell{{\sc{autostructure}}\\\citep{Badnell11}} & \makecell{{\sc{hullac}} \citep{BarShalom01}\\ {\sc{grasp2k}} \citep{Jonsson13}} & \makecell{Los Alamos Suite\\\citep{Fontes15b}} \\
Elements Used & \makecell{Z  $=$ 21$-$28 (d-block) \\ Z $=$ 58$-$70 (f-block)} & Se, Ru, Te, Ba, Nd, Er & Cr, Pd, Se, Te, Br, Zr, Sm, Ce, Nd, U\\
Opacity Method & Sobolev approx. & Sobolev approx.  & \makecell{Line-smearing approx.\\ \citep{Fontes15,Fontes17}} \\
Ejecta Modeled & \makecell{Grid of mass, velocity,\\and $\log\XLa$} & \makecell{Dynamical ejecta and two wind \\ compositions from \citet{Wanajo14}} & \makecell{Dynamical ejecta model \citep{Rosswog14}\\and two wind models \citep{Perego14}} \\
\enddata
\end{deluxetable*} 

The composition of the ejected material imprints itself on KN observations primarily through the radioactive heating rate and the ejecta opacity. The heating rate can be inferred from the decay-rate of the observed bolometric light curve. 
Nuclear network calculations show that the $r$-process will lead to a heating rate that converges to a power-law of the form \.{q} $\propto$ t$^{-1.3}$ \citep{Li98,Metzger10,Roberts11,Hotokezaka17,Rosswog17}.
This unique heating rate, due to the large number of radioactive isotopes in $r$-process material, is a clear signature of $r$-process synthesis, but it also unfortunately means that the heating rate contains little information about the \emph{detailed} composition at early times when the KN is brightest.
At later times, the heating rate could become dominated by a few individual isotopes, though the details are still quite uncertain \citep{Barnes16,Kasen18,Kasliwal18}.

Instead, the largest way the composition---in particular, the lanthanide fraction---of the ejecta will influence the observed properties of KN is though the \emph{opacity} of the material. Atomic structure models show that the lanthanides (with open valence electron f-shells) possess opacities $10-100{\times}$ larger than iron-like nuclei (with open d-shells) \citep[e.g.,][]{Kasen13, Tanaka13, Fontes17}. Qualitatively, the presence of a non-negligible amount of lanthanides will yield a fainter, redder, and longer lived kilonovae \citep[e.g.,][]{Barnes13}. Quantitatively, if modeling of the light curve of a KN requires an equivalent grey opacity larger than the $\kappa = 0.1 - 1$ cm$^2$ g$^{-1}$ expected for iron-like nuclei \citep{Pinto00}, it can be taken as evidence for lanthanide production.

However, linking KN observations to a \emph{specific} lanthanide fraction requires the determination of \emph{wavelength specific} opacities for different mixtures of $r$-process material. In practice, the atomic structure and line lists are woefully uncertain for these elements, so approximations must be made. This is also why detailed interpretation of composition from spectral features should currently be taken with caution.

Currently, three groups \citep[see][for the most recent publications]{Kasen17, Tanaka18, Wollaeger18} have computed numerical KN models with wavelength-specific opacities as an input for radiative transfer calculations, over a range of ejecta properties. These models can be fit to KN observations and constrain the ejecta lanthanide fraction. Each group utilizes a different radiative transfer code, theoretical atomic structure code, set of elements analyzed, and method for combining the atomic line data into wavelength-dependent opacities. We summarize these codes and model assumptions in Table~\ref{tab:kilonovamodels} and comment briefly on some salient features. 

The opacities used in radiative transfer are derived using theoretical atomic structure calculations of absorption lines, combining them into mean wavelength specific opacities including the effect of the significant ejecta velocity.
Not every model calculates the structure of all atoms, instead often using representative elements (e.g., Nd to represent most lanthanides).
\citet{Kasen17} and \citet{Tanaka18} use the Sobolev approximation to handle a large number of lines, while \citet{Wollaeger18} use a ``line-smearing'' approach that yields significantly larger opacities \citep{Fontes15,Fontes17}.
The models also differ in their treatment of the radioactive heating and thermalization, which affects the final inferred ejecta mass.

The physical properties of the ejecta components modeled by each group also differ. \citet{Kasen17} conside self-similar spherical ejecta, allowing them to consider a large grid in which the ejecta mass, velocity, and lanthanide fraction are all varied. In contrast, \citet{Tanaka18} and \citet{Wollaeger18} each present model light curves for three specific ejecta compositions, selected from numerical simulations to represent expectations for lanthanide-rich dynamical ejecta and two lanthanide-poor accretion disk wind models. Currently, only \citet{Wollaeger18} present models for asymmetric ejecta viewed at different angles.

\subsection{Composition constraints from GW170817}\label{sec:GW170817data}

\begin{figure}
    \centering
    \includegraphics[width=\linewidth]{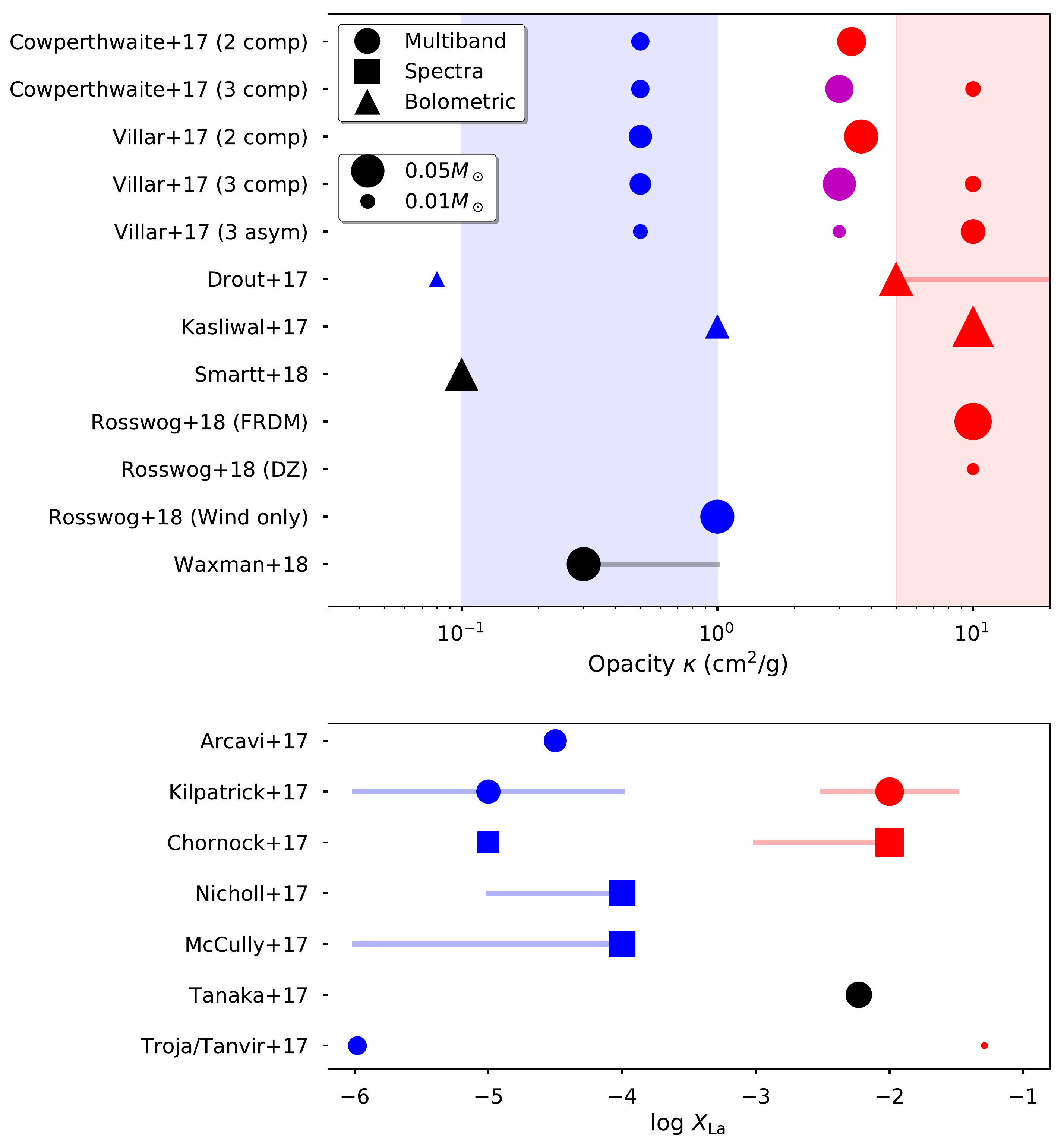}
    \caption{
    Literature compilation of inferences made for the composition of the G170817 kilonova.
    The top panel shows constraints on a grey opacity, where blue and red shaded regions indicate typical opacities for Fe-peak and lanthanide elements.
    The bottom panel shows constraints on the ejecta lanthanide fraction.
    The shape of each symbol shows what type of data was used to make the inference: a circle for multiband photometry, a square for an optical or infrared spectrum, and a triangle for the bolometric luminosity.
    In both panels, the area of each symbol is proportional to the mass in that component. 
    The symbol location is the most likely value, while the horizontal lines indicate uncertainties and/or a range of allowed values.
    When multiple ejecta components are modeled, different components are colored as blue, purple, or red. Black symbols indicate a single model for the entire light curve.
    Note that the total lanthanide fraction requires summing all mass components.
    }
    \label{fig:gw170817}
\end{figure}

GW170817 was the first stringent test of kilonova models.
Observations revealed thermal emission that rose on a timescale $\lesssim$12 hours, initially peaked in the UV, rapidly cooled from a temperature of $>$10,000 K to 3,000 K over 5 days, and was followed by a longer lived infrared transient (see Figure \ref{fig:gw170817-var}, left panel).
The early bluer ejecta and longer-lived redder ejecta had velocities of $\gtrsim$0.3c and $\sim$0.1c, respectively \citep{Shappee17,Chornock17}.
Overall, these observations closely matched theoretical expectations. It is broadly accepted that the observed transient is dominated by a KN and that a significant amount of $r$-process material was synthesized. 
Here, we summarize and compare works that modeled the UVOIR emission from GW170817 and offered \emph{quantitative assessments} of the ejecta composition. The results are plotted in Figure~\ref{fig:gw170817}.

\subsubsection{Grey opacity constraints}
The top panel of Figure~\ref{fig:gw170817} shows results from groups that constrain a grey opacity. Blue and red shaded regions indicate the equivalent grey opacities for iron-like ($\kappa \sim 0.1-1$ g cm$^{-2}$) and lanthanide-rich ($\kappa \sim 10$ g cm$^{-2}$) $r$-process ejecta. When groups required multiple ejecta components with different compositions to model GW170817, they are plotted on the same line, with symbol sizes scaled to inferred mass of each component. Different symbols indicate if models were made to reproduce multi-band or bolometric light curves. 

These grey opacity models do not directly provide a lanthanide fraction, but they generally require both lanthanide-poor \emph{and} lanthanide-rich material to be ejected.
The rapidly rising early blue ejecta is \emph{inconsistent} with the high-opacity lanthanide-rich ejecta, and is generally attributed to radioactive decay of light $r$-process elements\footnote{In the context of kilonovae, ``light $r$-process'' elements typically include elements in both the first and second peaks ($A{<}140$). This contrasts with metal-poor stars, where ``light $r$-process'' elements are typically restricted to just the first peak.}.
However, the presence of longer-lived infrared emission requires an additional component of lanthanide-rich ejecta.
Other lines of evidence for lanthanide-rich ejecta include the temperature plateau at 2500 K after ${\sim}5$ days \citep{Drout17,Kasliwal17}; near-infrared spectral features \citep{Chornock17,Pian17}; and the late-time bolometric light curve \citep{Kasliwal17,Kasliwal18,Villar18}.

Not all analyses arrived at this interpretation. Some groups argued that all the data could be reproduced with a single ejecta component and low opacity ($\kappa{\sim} 1$ g cm$^{-2}$, \citealt{Waxman18,Smartt17,Rosswog18}), but these models require fine-tuned compositions or thermalization efficiencies \citep{Kasen18}.
Other groups have argued that the early blue emission could be dominated by non-$r$-process energy sources, such as shock cooling \citep{Piro17}, ejecta from NS remnant winds \citep{Metzger18}, free-neutron decay \citep{Kasliwal17}, or long-term central engine activity \citep{Kisaka16}.
These energy sources would reduce (or even eliminate) the $r$-process mass inferred from the early blue component.

\subsubsection{Lanthanide fraction constraints}
In the lower panel of Figure~\ref{fig:gw170817}, we plot constraints on the \emph{lanthanide fraction}, X$_{\rm{La}}$, of the ejecta from GW170817 derived from fitting either observed light curves (circles) or spectra (squares) to the numerical KN models described above.  As in the top panel, when multiple components were required, they are plotted on the same line, with symbols scaled to the total mass of the component. Note that \citealt{Arcavi17}, \citealt{McCully17}, and \citealt{Nicholl17} \emph{only} model the early blue emission.
For fits performed using the models from \citet{Tanaka18} or \citet{Wollaeger18} we compute X$_{\rm{La}}$ from the detailed abundance patterns provided in those papers.

Each source here uses an independent data set, but not independent underlying models. 
\citet{Arcavi17}, \citet{Kilpatrick17}, \citet{Chornock17}, \citet{McCully17}, and \citet{Nicholl17} all utilize the grid of KN models provided by \citet{Kasen17}.
They all find that GW170817 is best reproduced by a combination of ${\sim}0.025 M_\odot$ of La-poor material at $v{\sim}0.3c$, and ${\sim}0.035 M_\odot$ of La-rich material at $v{\sim}0.1c$.
\citet{Tanaka17} use models from \citet{Tanaka18} and reproduce much of the emission of GW170817 with $\sim$0.03 M$_\odot$ of moderately lanthanide-enhanced $\log\XLa \sim -2.23$ material. 
They note there is room in their fits for sub-dominant ejecta components with varied compositions. 
\citet{Tanvir17} and \citet{Troja17} use the grid of models from \citet{Wollaeger18} and find a best fit to the emission from GW170817 with 0.015 M$_\odot$ of polar ``wind'' material with $\log\XLa = -6$ at $v \sim 0.08c$; and 0.002 M$_\odot$ of toroidal ``dynamical'' ejecta with $\log \XLa = -1.3$ and $v \sim 0.2c$, viewed at an angle of 30 degrees. We note that the ejecta velocity-composition pairings available in the publically released models of \citet{Wollaeger18} were based on numerical simulations, but are inconsistent with observations of GW170817---highlighting the need for an expanded model grid.

In order to arrive at a final X$_{\rm{La}}$ value for the total ejecta from GW170817, which can be compared to metal-poor stars, we combine together the contributions from multiple ejecta components when applicable. \emph{This inherently assumes that the early blue emission is predominately powered by the radioactive decay of light $r$-process elements}.
With this assumption, despite the differences in the numerical models used, the best fit models of \citet{Kilpatrick17}, \citet{Chornock17}, \citet{Tanaka17}, \citet{Tanvir17}, and \citet{Troja17} all produce a final, combined, lanthanide fraction of {\bf $\log \XLa = -2.2 \pm 0.5$}. Errors on this value include both composition ranges quoted as well as a factor of $\sim$3 error in the mass of each ejecta component, in order to account for systematic uncertainties in the thermalization efficiency, heating rate, and geometry \citep{Metzger17}.
Further systematics that could influence this derived $\XLa$ value are discussed in Section~\ref{s:compareKNstar}.

\subsection{Composition constraints from short-duration gamma-ray bursts}\label{s:sGRB}

While GW170817 provided the ``smoking gun'' observation that NSMs produce $r$-process nuclei, short-duration gamma-ray bursts (sGRBs) are also thought to originate from compact-object mergers. Thus optical/infrared observations of sGRBs can potentially provide constraints on the composition of the ejecta for more events. There have already been a number of claimed detections of KN associated with sGRBs \citep[e.g.,][]{Tanvir13,Berger13}. However, unlike the case of GW170817, sGRBs are viewed ``down the barrel'' of the relativistic jet. As a result, a substantial fraction of the observed optical/infrared light can be attributed to afterglow emission produced by the GRB jet \citep[e.g.,][]{Hjorth05}, making it difficult to disentangle potential KN emission. Furthermore, sGRBs are typically observed at high redshift and late-time infrared observations are quite limited. This will affect any inferred global lanthanide fraction. 

With all of these caveats in mind, \citet{Ascenzi18} recently re-examined all low redshift (z$<$0.5) sGRBs with detected optical or infrared afterglows. They model each event with a combination of jet afterglow and KN models, and use this to assess if there is any excess light which can be attributed to a KN. They use the numerical KN models from \citet{Kasen17}. In total, they identify three suspected KNe with with quite low $\log\XLa \lesssim -3$ and 1 KN (associated with GRB130603B) with $\log\XLa \sim -1.5$ to $-2.0$.

\section{Discussion}\label{s:discussion}
\subsection{Comparison of inferred KN ejecta to X$_{\rm{La}}$ distribution from metal-poor stars} \label{s:compareKNstar}

\begin{figure}
    \centering
    \includegraphics[width=\linewidth]{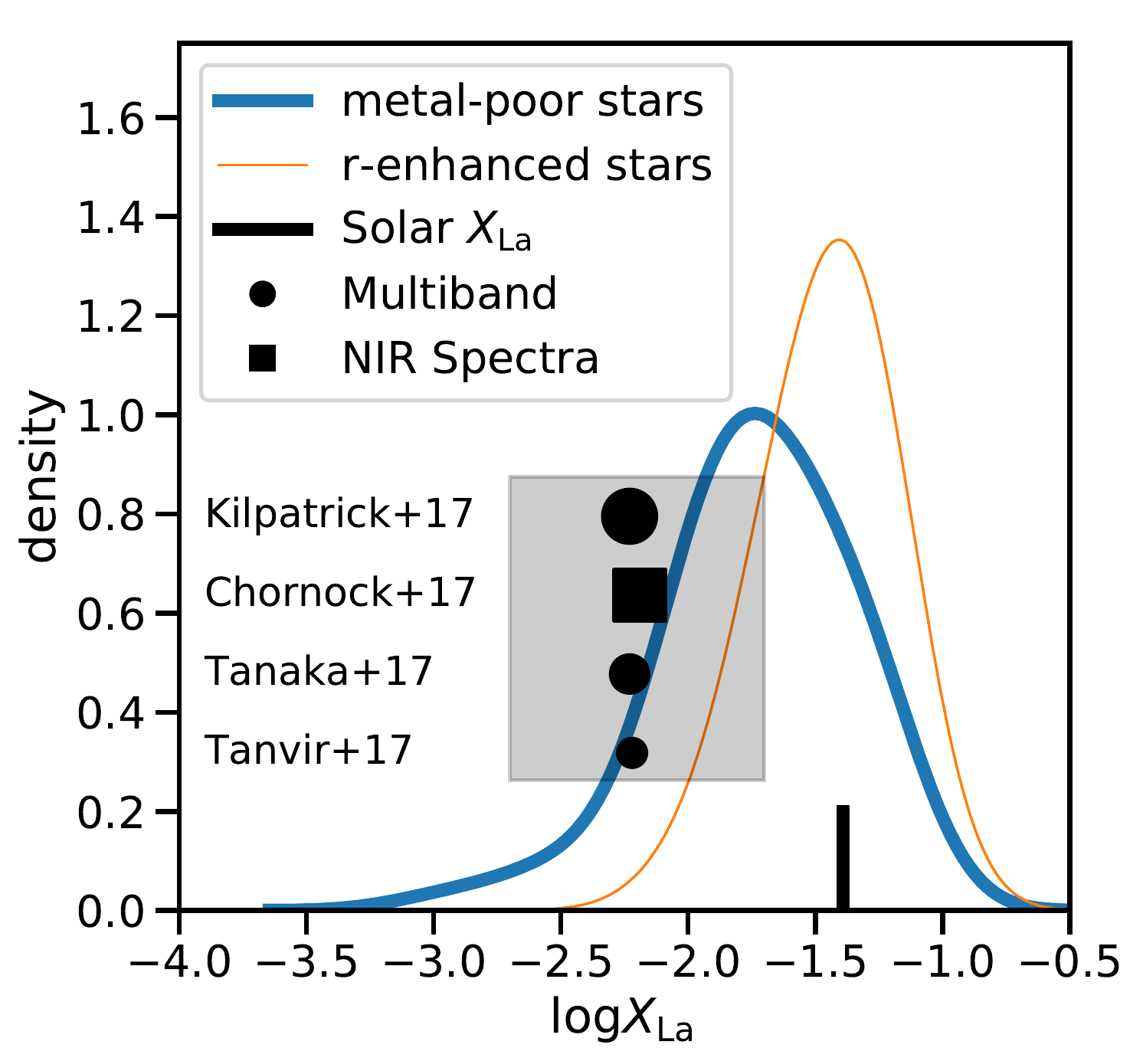}
    \caption{Summary of results. The thick blue line is the expected $\log\XLa$ distribution of $r$-process sites based on all metal-poor stars (JINAbase sample, A07 solar pattern). The thin orange line is the $\log\XLa$ distribution from the most $r$-process enhanced stars (R18 sample). 
    The short black line shows the Solar $\XLa$ value.
    The four black symbols indicate four independent determinations of the \emph{total} $\log\XLa$ from GW170817, derived by combining the mass and lanthanide fractions of all mass components. All four cluster closely around $\log\XLa \sim -2.2$, although the total ejecta masses differ by up to a factor of 3.5 ($0.017 M_\odot$ to $0.06 M_\odot$).
    The vertical position of these points is arbitrary.
    The grey shaded region represents the approximate error bar of $\pm 0.5$ dex for $\log\XLa$ in GW170817.
    About 10\% of future kilonovae should display substantially higher $\log\XLa > -1.5$.}
    \label{fig:summary}
\end{figure}

In Figure~\ref{fig:summary} we plot constraints on the total lanthanide faction of the GW170817 ejecta in comparison to the JINAbase sample of metal-poor stars (blue) and R18 sample of the most $r$-processed enhanced stars (orange). At log X$_{\rm{La}}$ $=$ $-$2.2 $\pm$0.5, the inferred value for for the ejecta of of GW170817 overlaps with the bulk of the JINA sample of metal-poor stars, but falls at the \emph{low end of the distribution}.
For the A07 (S08) solar abundance pattern, 82\% (60\%) of the JINA metal-poor stars have lanthanide fractions above log X$_{\rm{La}}$ $=$ $-$2.2. Notably, the X$_{\rm{La}}$ value for GW170817 appears as an outlier in comparison to the distribution $r$-enhanced stars from R18. For the A07 (S08) solar abundance pattern, 95\% (90\%) of R18 stars have lanthanide fractions above log X$_{\rm{La}}$ $=$ $-$2.2. In addition, while we caution that the lack of late-time infrared observations can preclude the identification of higher lanthanide fraction ejecta components, 3 of the 4 claimed KN detections from sGRBs in \citet{Ascenzi18} would also fall in the extreme low end of the X$_{\rm{La}}$ derived for metal-poor stars. 
Several other groups have also noted that $\XLa$ for GW170817 is low compared to the Sun \citep[e.g.,][]{Cote17}, but Figure~\ref{fig:summary} shows that will remain true for individual KNe.

We highlight three systematics that may influence our measurement of X$_{\rm{La}}$ from KN, impacting our comparison to the X$_{\rm{La}}$ distribution of metal poor stars. First, in deriving our final X$_{\rm{La}}$ for GW170817, we have assumed that the early blue component of the light curve is powered predominantly by the radioactive decay of light $r$-process elements. If other power sources contribute--or dominate--then the mass of light $r$-process elements produced may be smaller, and thus the total lanthanide fraction of the ejecta higher. For GW170817, we can examine the influence of this uncertainty by neglecting the contribution from the early blue component entirely. In particular, \citet{Kilpatrick17}, \citet{Chornock17}, and \citet{Kasen17} all find a moderately lanthanide enhanced composition for the \emph{red} component of GW170817 of $\log$ X$_{\rm{La}}$ $=$ $-2$ $\pm$ 0.5. This value falls closer to the median value for the bulk population, but still on the low end of the distribution for $r$-enhanced metal-poor stars: 63\% of the JINAbase and 94\% of the R18 samples have X$_{\rm{La}}$ values above that inferred for the dominant \emph{lanthanide-rich} ejecta of GW170817.

Second, it is possible that numerical models fit to the light curves and spectra of GW170817  over first $\sim$25 days post-merger do not fully capture all of the ejecta components. In particular, the high mass ($\sim$0.05 M$_\odot$) and low velocity ($\sim$0.1c) of the red component of GW170817 are \emph{inconsistent} with predictions from numerical simulations for the dynamical ejecta from NSMs---leading to the interpretation that it is actually produced by an accretion disk wind \citep[e.g.,][]{Metzger17}. However, tidally stripped dynamical ejecta should be ubiquitous in NSMs, and has been robustly predicted to be lanthanide-rich with $\log$ X$_{\rm{La}}$ $\approx$ $-1$ \citep[e.g.,][]{Roberts11,Rosswog14}. 

For the NS masses inferred from the gravitational wave signal of GW170817, numerical relativity simulations predict dynamical ejecta masses in the range of 10$^{-3}$ $-$ 10$^{-2}$ M$_\odot$ \citep{LIGODyn2017}. While the expected signal for this low-mass, high lanthanide fraction, ejecta component is faint, there may be evidence for it in current light curve fits. In particular, the three-component, grey opacity, models of \citet{Cowperthwaite17} and \citet{Villar17} favor the inclusion of a sub-dominant, high-opacity, ejecta component. However, we can conservatively estimate the impact of ``missing'' dynamical ejecta, by assuming the current models do not trace this component at all. The inclusion of an additional 10$^{-3}-10^{-2}$ M$_\odot$ of $\log$ X$_{\rm{La}}$ $=$ $-1$ material would raise the total X$_{\rm{La}}$ for GW170817 to $\log$ X$_{\rm{La}}$ $=$ $-1.9$ $\pm$ 0.5. 

Third, since KNe are powered by decaying $r$-process elements, a potential concern is that $\XLa$ will change between the initial composition measured in KNe and the several-Gyr old composition measured in metal-poor stars.
Fortunately, it turns out that the KN composition from the 2nd-to-3rd peak does not change significantly after 1 day \citep[see fig 7 in][]{MendozaTemis15}.
However, the actinides have many longer-lived isotopes, and their abundance \emph{does} change significantly between 1 day and 1 Gyr.
We have neglected actinides in this paper because they contribute negligible mass in metal-poor stars and the Sun, but actinides could affect $\XLa$ measurements from KNe.
If this effect is important, then the very neutron-richest ejecta ($Y_e < 0.1$, \citealt{Holmbeck19}) could cause $\XLa$ variations in KNe that are not currently constrained by metal-poor stars: the only remaining actinides are the long-lived but still radioactive isotopes of Th and U, and long-lived actinide decay products like Pb and Bi are rarely (if ever) measured in stars. 
Unfortunately, it also is known that the relative actinide abundance varies significantly in $r$-enhanced metal-poor stars \citep[e.g.,][]{Hill02,Ren12,Mashonkina14,Ji18,Holmbeck18,Holmbeck19}. The topic of actinides in both stars and KNe clearly deserves further study (see, e.g., \citealt{Holmbeck19b}). We note that the choice made here to ignore actinides means that all our star $\XLa$ constraints could be considered lower limits when comparing to KNe, which would increase the mild tension between GW170817 and most stars.

\subsection{Implications for future KN observations}\label{s:KNobs}

We find that estimates for the total X$_{\rm{La}}$ of GW170817 fall on the low end of the distribution observed in metal-poor stars. 
Thus, \emph{if NSMs are the dominant $r$-process source, we should expect future observations to reveal a significant number of KN with higher overall lanthanide fractions than that observed in GW170817}. As the total lanthanide fraction produced in the ejecta of a given NSM is likely due to a combination of different ejecta components with different compositions, the requirement for a higher total lanthanide fraction could manifest itself on KN observations in multiple different ways. Here we describe the baseline level of observations required to constrain the overall lanthanide fraction of a KN, as well as \emph{merger model-dependent} implications for future observations.  

\begin{figure}[th]
\centering
\includegraphics[trim={1.3cm 0.7cm 0.9cm 0.2cm},clip,width=0.93\columnwidth]{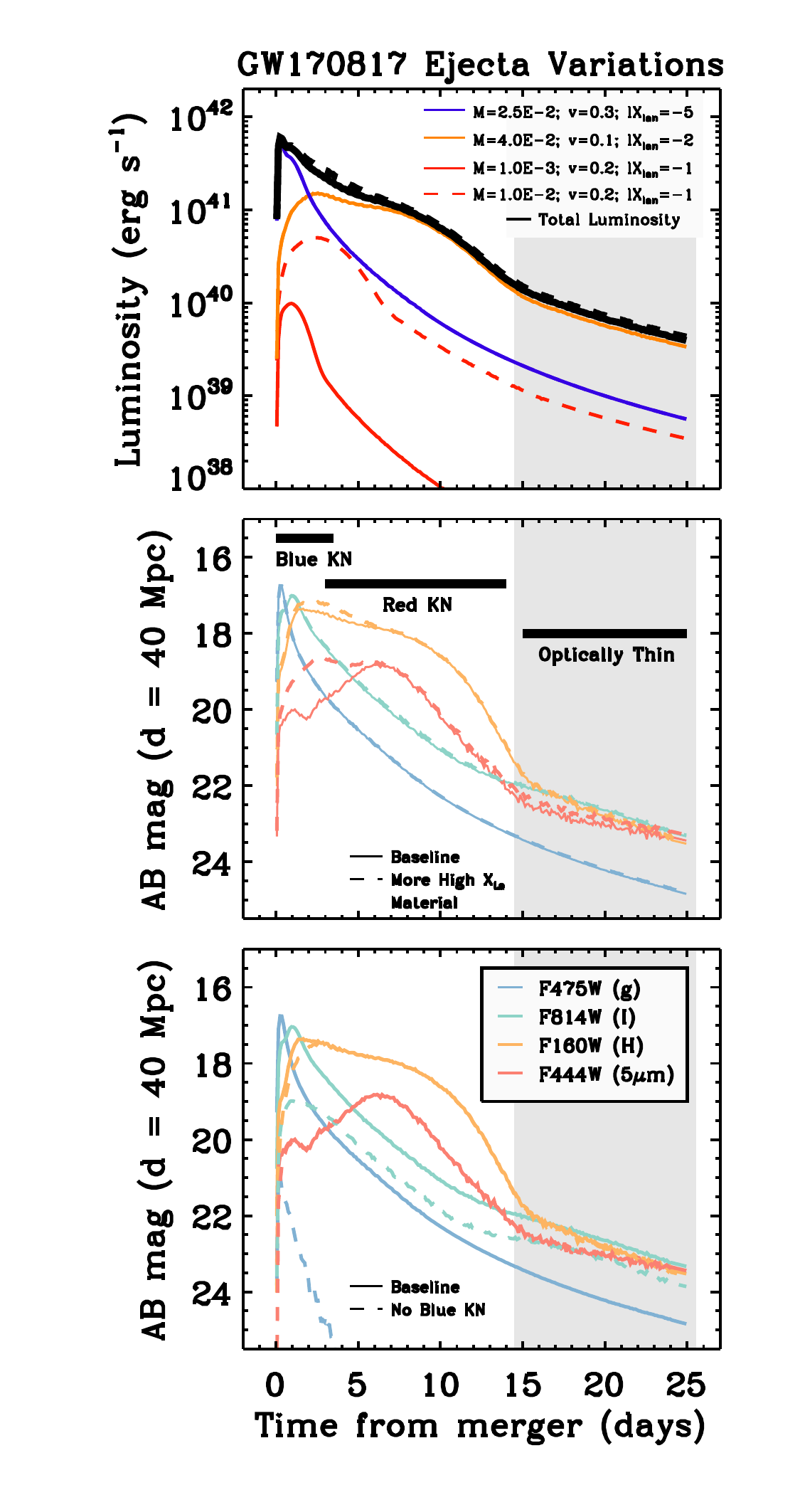}
\caption{
\emph{Top:} Bolometric luminosity from individual NSM ejecta components (colored lines). Solid lines represent a baseline 3-component model for GW170817. Dashed lines show the affect of an enhanced mass of very $\XLa$-rich material. Models from \citet{Kasen17}. See text for details.
\emph{Middle:} Light curves associated with the baseline (solid lines) and $\XLa$-rich (dashed lines) 3-component models from the top panel. Increasing the mass of $\XLa$-rich material in GW170817 from 10$^{-3}$ to 10$^{-2}$ M$_\odot$ would be observable primarily in the mid-infrared light curves.
\emph{Bottom:} Light curves comparing the baseline model (solid lines) to a model with no blue KN component (dashed lines).
\label{fig:gw170817-var}}
\end{figure}

First, 
robustly constraining the total lanthanide fraction from a given KN requires \emph{both} UV/blue and (mid-)infrared observations over the first $\sim$week post-merger, as they predominantly probe lanthanide-poor and lanthanide-rich ejecta, respectively. This is demonstrated in Figure~\ref{fig:gw170817-var} where we highlight the impact of slight variants on the ejecta from GW170817 to the observed light curves. In the top panel we use the models of \citet{Kasen17} to plot the bolometric light curves for four fiducial ejecta components. We plot a lanthanide-poor ``blue kilonova'' and a moderately lanthanide-enhanced ``red kilonova''---both with masses, velocities and lanthanide-fractions consistent with those inferred for GW170817. These are supplemented by two different masses of lanthanide-rich ($\log\XLa = -1$) ejecta, meant to represent a range of possibly unaccounted for high $\XLa$ material (see above). 

In the middle and lower panels, the solid lines represent the combined $g-$, $I-$, $H-$, and 5 $\mu$m light curves for a ``baseline'' GW170817 model of the blue KN, red KN, and $10^{-3}$ M$_\odot$ of high $\XLa$ ejecta. In the middle panel, dashed lines demonstrate the effect of increasing the mass of the lanthanide-rich ejecta to $10^{-2}$ M$_\odot$, which would increase the overall $\log \XLa$ from $-2.2$ to $-1.9$. The difference is primarily discernible in light curves 5$\mu$m and redward, highlighting the importance of space-based mid-infrared observatories, such as Spitzer and JWST to constrain the presence of lanthanide-rich components. Overall, neglecting any infrared observations (at $\lambda {>}1\mu m$) would cause grossly underestimated lanthanide fractions ($0.5-1$ dex in these models). In addition, while the models of \citet{Kasen17} break down in the optically thin phase (shaded region), observations on these timescales may also be sensitive to the relative abundance of light and heavy $r$-process elements \citep{Kasen15,Kasen17}. 

\begin{figure*}[t]
\centering
\includegraphics[trim={1.3cm 0.8cm 1cm 0},clip,width=\linewidth]{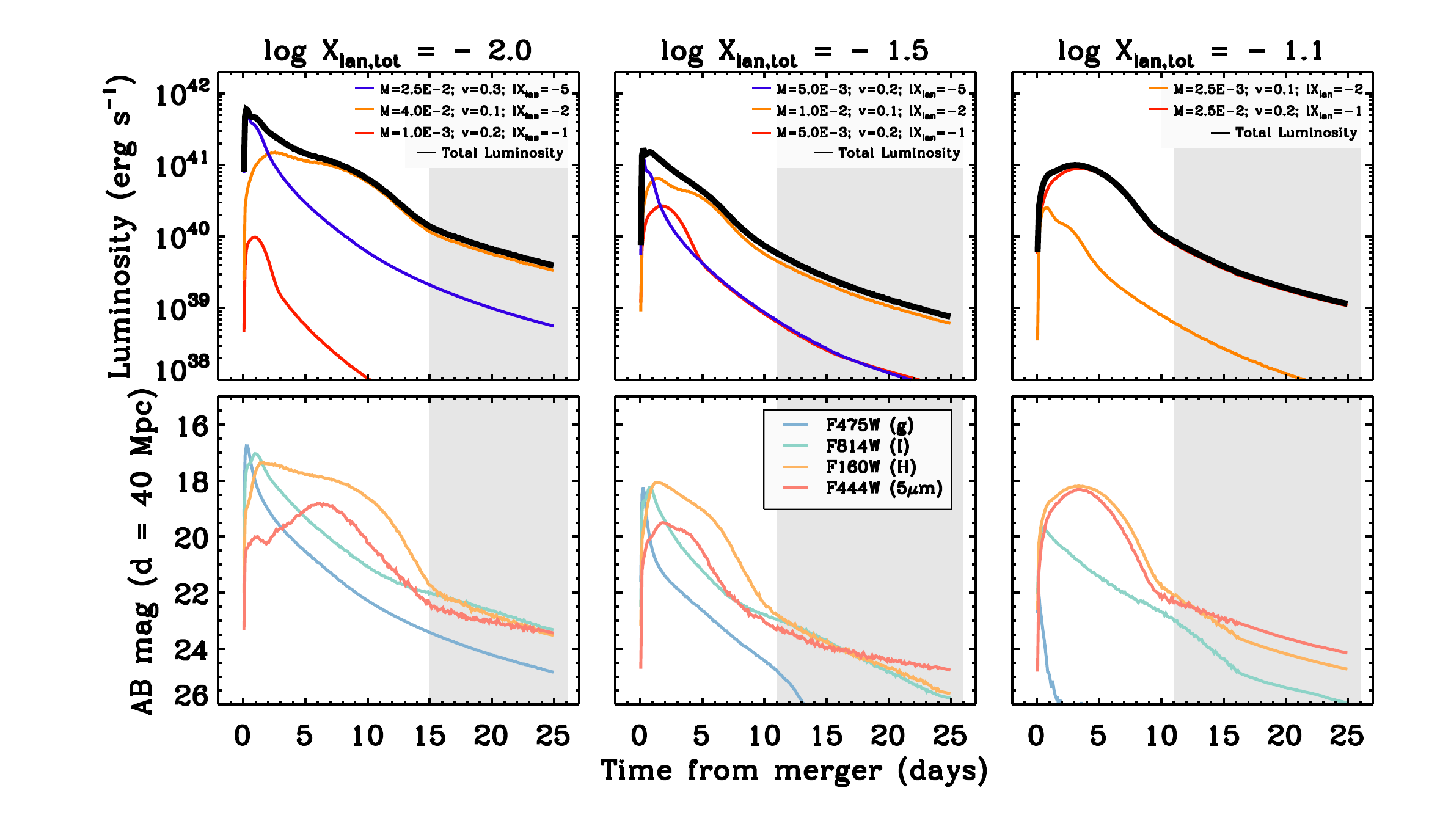}
\caption{Representative examples of expectations for higher-$\XLa$ kilonovae.
The top row shows bolometric light curves. Red, blue, orange, and black lines represent models for $\XLa$-rich tidal dynamical ejecta, $\XLa$-poor shocked dynamical ejecta, moderately $\XLa$-enhanced accretion disk winds, and total combined luminosity, respectively. The lower panels show corresponding light curves for each combined model in four filters. 
\emph{Left column:} baseline GW170817 model with $\XLa \sim 10^{-2}$ (same as top two panels of Figure~\ref{fig:gw170817-var}).
\emph{Middle column:} $\XLa \sim 10^{-1.5}$ model, achieved by lowering the mass ejected in disk winds as in prompt BH formation.
\emph{Right column:} $\XLa \sim 10^{-1.1}$ model, achieved by having ejecta dominated by a tidal tail as in a NS-BH merger.
All models from \citet{Kasen17}. See text for details.
\label{f:observed_kn}}
\end{figure*}

In the lower panel, the dashed lines demonstrate the effect of removing the blue KN ejecta component from GW170817, which would increase the overall $\log \XLa$ from $-2.2$ to $-2.0$. This demonstrates the necessity of blue/UV observations to probe the presence/lack of lanthanide-poor ejecta components. A search done entirely in the infrared would cause total lanthanide fractions to be overestimated. In particular, while blue light dominated IR for only $\sim$ 2 days in GW170817, and UV observations in the first $\lesssim$ 12 hours post-merger may be particularly valuable in discerning the \emph{origin} of early blue emission \citep{Arcavi18}, blue observations at later times are still useful for discerning the presence of lanthanide-poor material. 

Figure~\ref{fig:gw170817-var} highlights the need for observations across the entire UVOIR range to constrain the total X$_{\rm{La}}$ from a NSM. However, it is difficult to directly translate the expectation of higher $\XLa$ into predictions for other observed properties such as timescale and luminosity in a \emph{model-independent} way, because the ejecta mass and velocity also play a significant role in determining these properties\footnote{Note that the diffusion time, $t_{\rm diff}$, has the following dependency: $t_{\rm diff} \propto M \kappa v^{-1}$; \citealt{Metzger17LRR}.}. We therefore rely on lessons from current relativistic hydrodynamic NSM models and GW170817 which suggest their should be 3 primary ejecta components: (i) fast and lanthanide-rich (log $\XLa$ $\sim$ $-$1) material dynamically ejected in tidal tails, (ii) fast and lanthanide-poor (log $\XLa$ $\sim$ $-$5) material---either due to material dynamically ejected from from NS interface (``shocked dynamical'') or produced by a wind from a remnant NS surface, and (iii) slower material produced by post-merger accretion disk winds, with a variable composition. In GW170817, the presumed post-merger accretion disk wind dominated the total ejected mass ($\sim$0.04 M$_\odot$) and possessed a lanthanide fraction of $\log\XLa = -2$ \citep{Siegel18}. Thus, in this context, there are two possible ways to increase the total $\XLa$ from a NSM relative to GW170817: to either change the masses of the three components such that the relative contribution from the high-$\XLa$ tidal dynamical ejecta is enhanced, or to increase the overall $\XLa$ of the accretion disk winds.

Current numerical simulations find that the contribution from tidally ejected material is enhanced relative to the shocked dynamical ejecta for cases of highly asymmetric binary mass ratios \citep{Bauswein13,Hotokezaka13,Dietrich2015,Sekiguchi2016}, binaries which undergo prompt ($\lesssim$1 ms) collapse to a black hole \citep{Radice2018}, and highly eccentric mergers \citep{Chaurasia2018}. Even in these scenarios, however, the total predicted tidal tail masses are small, ranging from 10$^{-4}$ $-$ 10$^{-2}$ M$_\odot$. Thus, significantly enhancing the overall contribution from the dynamical ejecta \emph{requires} a smaller disk wind mass (and hence smaller total ejecta mass) than observed in GW170817. Physically, numerical simulations find that the total disk mass will be suppressed for low values of the tidal deformability of the binary, $\Tilde{\Lambda}$, typically corresponding to prompt black hole (BH) formation \citep{Radice2018,Radice2018GW}. 

In addition, while in principle the total lanthanide fraction of the a NSM ejecta could be increased by increasing $\XLa$ of the disk wind---rather than decreasing its mass---in practice higher $\XLa$ disk winds are typically \emph{linked} to lower total disk wind masses in merger models. This is because higher $\XLa$ accretion disk winds are primarily achieved by reducing the exposure of the disk to protonizing neutrinos from a central post-merger NS remnant. The dominant factor influencing this exposure is thus the lifetime of any post-merger NS remnant \citep[e.g.,][]{Lippuner17}. Prompt BH formation will yield both lower mass and higher $\XLa$ disk winds. Alternatively, neutrino exposure can also be reduced (and $\XLa$ increased) by ejecting disk material on shorter timescales \citep[e.g.][]{Wu16}, which also tends to reduce the total ejecta mass.    

Thus, we find that \emph{in the context of current merger models} achieving a significantly higher $\XLa$ than that observed in GW170817 \emph{typically requires smaller total ejecta masses}. As a result, if NSMs are the dominant source of the $r$-process, \emph{many future KN should be fainter, redder and more rapidly-evolving} (and thus more difficult to identify and follow) than GW170817. This is demonstrated in Figure~\ref{f:observed_kn}. In the left panels we again plot our baseline three-component model for GW170817, with a total $\log\XLa = -2.0$. In the middle panels we show a \emph{representative example} of merger which ejects 5$\times$10$^{-3}$ M$_\odot$ of lanthanide-rich tidal dynamical and lanthanide-poor squeezed dynamical ejecta, and 10$^{-2}$ M$_\odot$ of a moderately lanthanide-enhanced disk wind, yielding a total ejecta lanthanide-fraction of $\log \XLa = -1.5$. This disk wind mass falls on the high end of what may be expected for the case of a prompt collapse to a BH in the models of \citet{Radice2018}, and already the resulting KN peaks more than a magnitude fainter and decays roughly a factor of 2 more rapidly than GW170817.

In addition, in comparison to binary NS mergers, NS-BH mergers can produce ejecta with higher total $\XLa$ values. In particular, the tidal dynamical ejecta is expected to dominate for multiple reasons. First, there is neither a merger interface surface to produce blue shocked dynamical ejecta nor a central NS remnant from which blue winds can be driven. Second, due to the large binary mass ratios, the total dynamical ejecta masses tend to exceed those achieved in binary NS mergers, reaching values of 1-4 $\times$10$^{-2}$ \citep{Foucart17}. Finally, although the exact mass and composition of the disk winds from NS-BH mergers is still rather uncertain \citep[e.g.,][]{Just15,Foucart17,Foucart18}, the ratio of the dynamical mass eject to post-merger accretion disk winds is probably higher in NS-BH mergers on average compared to NS-NS mergers (see fig~1 in \citealt{Wu16}). In the right panels of Figure~\ref{f:observed_kn} we use the models from \citet{Kasen17} to plot a representative example ejecta that may result from a NS-BH merger: 2.5$\times$10$^{-2}$ M$_\odot$ of lanthanide-rich tidal dynamical ejecta and 2.5$\times$10$^{-3}$ M$_\odot$ moderately-enhanced disk wind, yielding a total $\log \XLa = -1.1$. Due to the lack of a lanthanide-poor ejecta component, the observed colors are significantly redder at all times.

Finally, we note there have been some suggestions that the most $r$-enhanced (and highest-$\XLa$) stars tend to occur at lower [Fe/H] \citep{Barklem05,Roederer18a,Brauer19}. Assuming [Fe/H] corresponds to time, this suggests a potential association between the merger delay time and the lanthanide fraction.
Observationally, that would imply that high-$\XLa$ KNe are more likely to be found in star-forming galaxies, where the typical NSM should have a shorter delay time (in contrast to GW170817, which is in one of the oldest host galaxies with a short gamma ray burst (GRB); \citealt{Fong17}).
However, while we find evidence that $\log\XLa$ is anticorrelated with [Fe/H] in the R18 sample, this anticorrelation is not significant in the full JINAbase sample.

\subsection{Implications for the origin of $r$-process elements}\label{s:discussionorigin}
Figure~\ref{fig:summary} shows that the observed $\XLa \sim -2.2$ for GW170817 is reasonably consistent with the bulk of metal-poor stars.
The most $r$-enhanced stars have higher $\XLa \gtrsim -1.5$, so we expect to find KNe with high $\XLa$ in future followup of NSMs.
Metal-poor star statistics suggest that ${\sim}10$\% of NSMs should have $\log\XLa > -1.5$, with most of the rest having $\log\XLa \sim -1.8$ (Figure~\ref{f:histogram_samples}).
Thus, followup of binary neutron star merger triggers from LIGO should be able to test consistency between these distributions within the next few years.

However, if future KNe followup of gravitational wave sources does \emph{not} find a high-$\XLa$ population, this would imply that metal-poor stars with the highest $r$-process enhancements are \emph{not} enriched by NSMs.
Several types of rare supernovae provide promising (but still theoretical) sources of $r$-process elements, such as the magnetorotationally driven jet supernova \citep[e.g.,][]{Winteler12,Nishimura15,Moesta18}, common envelope jet supernovae \citep{Grichener18}, or collapsar disk winds \citep[e.g.][]{Surman06,Siegel18}.

In fact, chemical evolution arguments have tended to prefer rare supernovae as the dominant source of $r$-process elements, at least in the early universe when metal-poor stars are forming \citep[e.g.,][]{McWilliam97,Argast04,Qian08,Cescutti15,Wehmeyer15,Cote18}.
Such models tend not to fully account for hierarchical galaxy formation, inhomogeneous metal mixing, or NSM velocity kicks; and including these effects can significantly affect chemical evolution interpretations \citep[e.g.,][]{Tsujimoto14b,Hirai15,Ishimaru15,Shen15,vandeVoort15,Ji16b,Bramante16,Beniamini16,Beniamini18a,Duggan18,Naiman18,Safarzadeh18b,Schonrich19}.
In light of these complications, we suggest that the composition of future KNe provides a more direct test of whether NSM ejecta are viable sources of the high $\XLa$ found in the most $r$-enhanced stars.

Many chemical evolution models have suggested that the dominant astrophysical site changes over time, with an early contribution from supernovae followed by subsequent enrichment by NSMs \citep[e.g.,][]{Cescutti15,Wehmeyer15,Cote18,Schonrich19}.
As metal-poor stars likely form at redshifts $z \gg 2$, and neutron star mergers detected in gravitational waves will be at $z \ll 1$, a mismatch in the $\XLa$ distributions might indicate time evolution in the dominant $r$-process site.

\subsection{Comparison to theoretical $\XLa$ from models of different astrophysical sites}\label{s:discussionmodels}
\begin{figure}[b]
\centering
\includegraphics[width=1\linewidth]{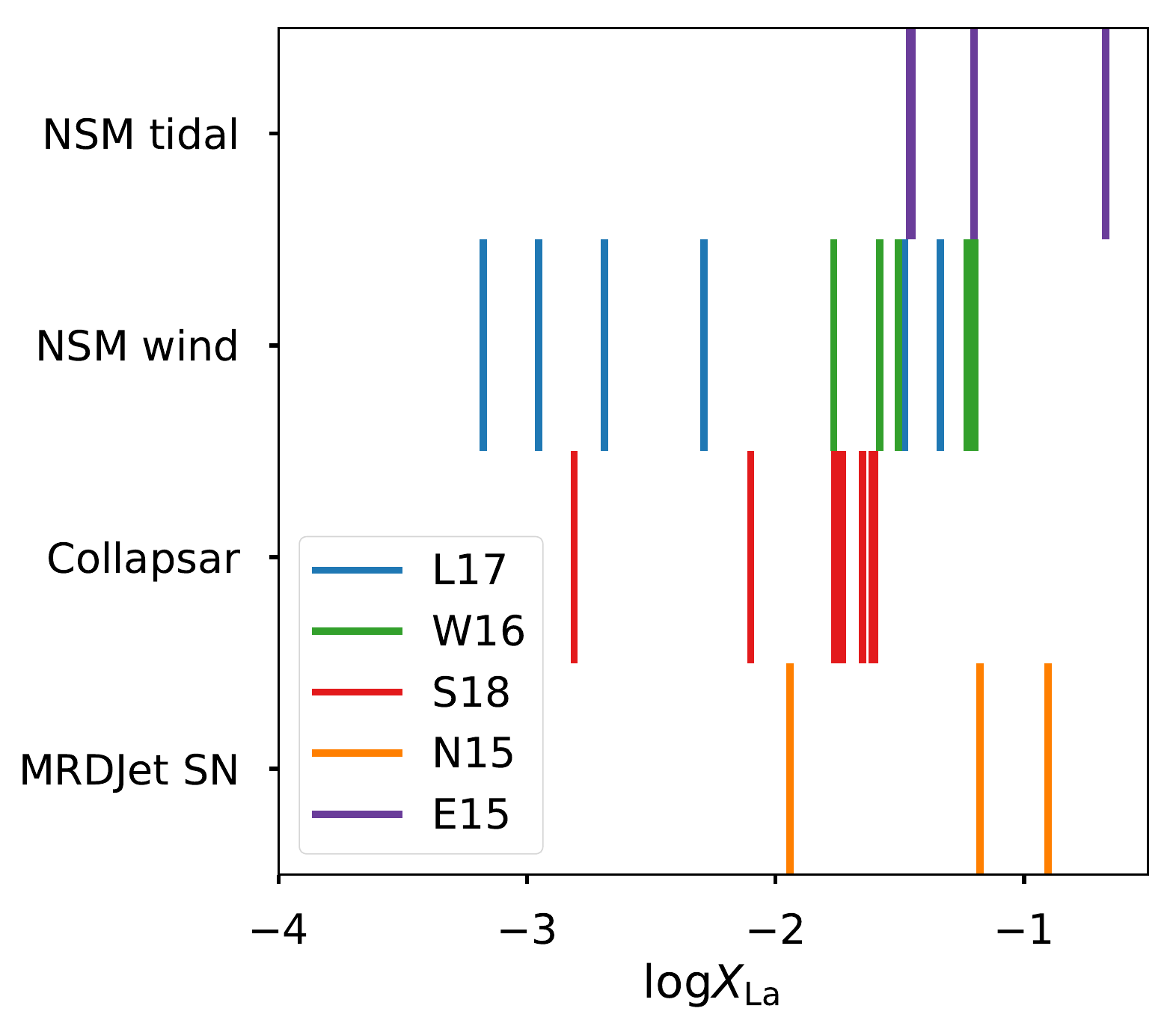}
\caption{Modeled $\log\XLa$ for three different astrophysical sites: NSM tidal ejecta \citep{Eichler15}, NSM disk wind \citep{Wu16,Lippuner17}, collapsar disk wind \citep{Siegel18}, and magnetorotationally driven jet \citep{Nishimura15}. Data in Table~\ref{tab:modelXLa}.
\label{f:modelXLa}}
\end{figure}

For theoretical context, we have compiled lanthanide fractions from several models of $r$-process sites and plotted them in Figure~\ref{f:modelXLa}.
We include models for neutron star merger tidal ejecta \citep[][E15]{Eichler15}, neutron star merger disk winds \citep[][L17, W16]{Lippuner17,Wu16},
collapsar disk winds \citep[][S18]{Siegel18}, and magnetorotationally driven jets \citep[][N15]{Nishimura15}.
We warn the reader not to make detailed comparisons between Figure~\ref{f:modelXLa} and the stellar lanthanide fraction distribution (Figures~\ref{f:histogram_samples} and \ref{fig:summary}).
There are still significant uncertainties with the nuclear data used in these calculations, and the theoretical $r$-process patterns deviate in quantitatively important ways from the solar $r$-process pattern, including near the lanthanide region \citep[e.g.,][]{Eichler15,MendozaTemis15,Barnes16,Mumpower16,Nishimura16,Wu16,Shibagaki16,Ji18}.
Furthermore, these models compute the yields from just a single ejecta component, and true astrophysical sites are a sum of many components (e.g., dynamical ejecta for the NSMs; neutrino-driven winds in the supernovae; a combination of both collapsar disk winds and MRD jets).

However, there are still two interesting points to take away. The first point is that all models here have a subset of parameter space that allows them to reproduce the highest $\XLa$ observed in metal-poor stars ($\XLa \gtrsim 10^{-1.5}$).
NSM tidal ejecta are naturally high $\XLa$.
In NSM disk winds, high $\XLa$ occurs when the disk experiences relatively low neutrino irradiation (e.g., with a short hypermassive neutron star lifetime; \citealt{Wu16,Lippuner17}).
The collapsar models produce high $\XLa$ when the mass accretion rate onto the black hole is high \citep{Surman06, Siegel18}.
And the MRD Jet SNe produce high $\XLa$ when the pre-collapse magnetic field is strong enough to promptly launch neutron-rich jets \citep{Nishimura15, Moesta18}.
The second point is that as the models develop, the observed stellar $\XLa$ distributions should provide interesting constraints on the physics of $r$-process sites.
For example, if NSMs dominate $r$-process production, the $\XLa$ distributions may provide independent information about the distributions of binary mass ratios, eccentricities, and neutron star equation of state \citep[e.g.,][]{Radice16, Lippuner17}.

Note that in this paper, our definition of lanthanide fraction is lanthanide mass over \emph{neutron-capture element mass}.
We do not use the total mass, because any other element would be significantly contaminated by core-collapse supernovae in metal-poor stars.
This is not a concern if neutron star mergers are the dominant $r$-process site, because negligible amounts of Fe are produced in those mergers.
However, if lanthanide fractions are eventually derived in other potential $r$-process sites like collapsars \citep{Siegel18}, one will have to remove the mass contribution of any Fe-peak elements in order to compare to metal-poor stars.

\section{Conclusion}\label{s:conclusion}

The dominant $r$-process site in the universe is still uncertain, but NSMs are a plausible source to explain $r$-process element production over all cosmic time.
In this paper, we show that comparing \emph{populations} of NSM kilonovae to \emph{populations} of metal-poor stars with $r$-process elements provides a new way to understand the origin of the $r$-process in this new multimessenger era.
We have provided half of the required data for this comparison, by computing the lanthanide fraction distribution from metal-poor stars.
The other half will come in time, as LIGO continues to detect more neutron star mergers in O3 and beyond.
One key result is that the GW170817 kilonova is rather lanthanide-poor compared to the bulk of metal-poor stars (Figure~\ref{fig:summary}).
\emph{If neutron star mergers are the dominant $r$-process site, we should expect most future kilonovae to be more lanthanide-rich, with 10\% having much higher lanthanide fractions $\log \XLa > -1.5$.}
Properly measuring kilonova lanthanide fractions requires observing both optical and (mid-)infrared light curves, in order to constrain the mass and lanthanide fraction of all ejected components.

We view this paper as a first step towards combining observations of KNe and metal-poor stars to understand the $r$-process. Throughout this paper, we have emphasized the many systematics of determining $\XLa$ in both these systems.
Significant theoretical advances still need to be made for $r$-process nucleosynthesis calculations and interpreting KNe compositions from observations.
These will be greatly helped by UV spectroscopy to observe abundances of currently extrapolated elements in stars, and also by observing more KNe.

We expect that the comparison between metal-poor star and KN lanthanide fractions will become statistically interesting once ${\sim}10$ KNe are followed up. For optimistic merger rates, this could happen as soon as the end of LIGO O3, though more likely it will take ${\sim}5-10$ years \citep[e.g.,][]{Cowperthwaite19}.
After this period, if high-$\XLa$ kilonovae are not detected, we regard this as very strong evidence for the existence of at least one additional $r$-process source that dominates $r$-process production in the early universe.

\acknowledgments
We thank Jennifer Barnes, Ryan Foley, Wen-Fai Fong, Kenta Hotokezaka, Charlie Kilpatrick, Jonas Lippuner, Tony Piro, Enrico Ramirez-Ruiz, Daniel Siegel, Nicole Vassh, and Salvo Vitale for useful discussions.
We thank Daniel Siegel, Meng-Ru Wu, and Marius Eichler for sharing their model yields.
We thank the Carnegie Observatories for hosting a vibrant postdoc community that helped spur this work.
Support for this work was provided to APJ and MRD through Hubble Fellowship grants HST-HF2-51393 and NSG-HF2-51373, respectively, both awarded by the Space Telescope Science Institute, which is operated by the Association of Universities for Research in Astronomy, Inc., for NASA, under contract NAS5-26555. MRD acknowledges support from the Dunlap Institute at the University of Toronto and the Canadian Institute for Advanced Research. 
This research has made use of the SIMBAD database, operated at CDS, Strasbourg, France \citep{Simbad},
and NASA's Astrophysics Data System Bibliographic Services.

\software{Astropy \citep{astropy}, numpy \citep{numpy}, matplotlib \citep{matplotlib}, seaborn \citep{seaborn}, pandas \citep{pandas}}

\appendix
\section{Literature Abundance References}\label{litrefs}
Stellar abundances from \citet{Roederer18a} compilation are sourced from:
\citet{Aoki10,Barklem05,Cohen13,Cowan02,Frebel07b,Hansen18,Lai08,Francois07,Hayek09,Hill02,Hill17,Hollek11,Holmbeck18,Honda04,Howes16,Ivans06,Jacobson15,Johnson13,Mashonkina10,Mashonkina14,Placco17,Roederer14c,Roederer18b,Sakari18,Sneden03,Westin00}.

Stellar abundances from \citet{jinabase} (JINAbase) compilation are sourced from:
\citet{Allen12,Aoki05,Aoki10,Barklem05,Cayrel04,Christlieb04,Cohen13,Frebel07b,Hansen15b,Hayek09,Hollek11,Honda04,Honda11,Ishigaki13,Jacobson15,Johnson02a,Lai08,Li15a,Li15b,Mashonkina10,Mashonkina14,McWilliam95,Placco14,Preston06,Roederer10,Roederer14c,SiqueiraMello14,Sneden03,Westin00}.

\section{Star Data Table}\label{xlatable}
\begin{table}[h!]
\label{tab:stardata}
\centering
\begin{tabular}{cccccccccc}
Star & Reference & Nelems & [Fe/H] & [Sr/Fe] & [Ba/Fe] & [Eu/Fe] & logXLa\_A07 & logXLa\_S08 & Sample \\
\hline
CS22882-001 & ROE14 & 9 & -2.62 & 0.16 & 0.06 & 0.81 & -1.51 & -1.81 & JINAbase \\
J235718-005247 & AOK10 & 7 & -3.36 & 0.78 & 1.08 & 1.92 & -1.15 & -1.12 & R18 \\
J0030-1007 & SAK18 & 3 & -2.35 & 0.50 & -0.71 & 0.00 & -2.69 & -2.93 & S18 \\
J00002259-1302275 & HAN18 & 3 & -2.90 & -1.20 & -0.38 & 0.58 & -1.01 & -1.17 & H18 \\
\end{tabular}
\caption{Star Data Table}
\end{table}

This table contains inferred lanthanide fractions for the stars studied in this paper.
We show four random stars, one from each source. A full table is available online.
Stars are duplicated when in multiple sources.

\newpage
\section{Model Data Table}\label{modeldatatable}
\begin{table}[h]
    \centering
    \begin{tabular}{cccc}
System & Model Number & Reference & $\log\XLa$ \\
\hline
Collapsar Wind & E15 & S18 & $-1.65$ \\
Collapsar Wind & E20 & S18 & $-1.61$ \\
Collapsar Wind & E25 & S18 & $-1.60$ \\
Collapsar Wind & E15B & S18 & $-2.81$ \\
Collapsar Wind & E20B & S18 & $-1.76$ \\
Collapsar Wind & F15B & S18 & $-2.10$ \\
Collapsar Wind & F20B & S18 & $-1.73$ \\
MRD Jet SN & B11W025 & N15 & $-\infty$ \\
MRD Jet SN & B11W100 & N15 & $-0.90$ \\
MRD Jet SN & B12W025 & N15 & $-1.94$ \\
MRD Jet SN & B12W100 & N15 & $-1.18$ \\
MRD Jet SN & B12W400 & N15 & $-0.77$ \\
NSM wind & Hinf & L17 & $-3.18$ \\
NSM wind & H300 & L17 & $-2.95$ \\
NSM wind & H100 & L17 & $-2.69$ \\
NSM wind & H030 & L17 & $-2.29$ \\
NSM wind & H010 & L17 & $-1.48$ \\
NSM wind & H000 & L17 & $-1.34$ \\
NSM wind & Sdef & W16 & $-1.50$ \\
NSM wind & m01 & W16 & $-1.20$ \\
NSM wind & m10 & W16 & $-1.58$ \\
NSM wind & s10 & W16 & $-1.77$ \\
NSM wind & s6 & W16 & $-1.23$ \\
NSM tidal & P01 & E15 & $-1.46$ \\
NSM tidal & KT75 & E15 & $-0.67$ \\
NSM tidal & P08 & E15 & $-1.45$ \\
NSM tidal & ABLA07 & E15 & $-1.20$
\end{tabular}
    \caption{$\XLa$ for different astrophysical $r$-process sites.
    The yields are reported in slightly different formats, so we describe the $\XLa$ calculation in more detail.
    L17 reports their average lanthanide fractions in their table~2.
    For N15, W16, and E15, we calculate $M_A$, $M_B$, and $M_C$ directly from the abundance patterns predicted for the models with the atomic mass ranges in Table~\ref{tab:categories} and compute $M_C$/($M_A+M_B+M_C$).
    For S18, abundance patterns were calculated at three specific points in an individual event, finding $\log M_{\rm La}/M_{\rm n-cap} \approx -1.6 $ for their $\dot M_1$ and $\dot M_2$ while $M_{\rm La}/M_{\rm n-cap} \approx 0$ for all other mass components (D. Siegel, private communication). Using their extended data table 1, we assume $\log\XLa = -1.6$ for the $r$-proc mass component, all of the light $r$-proc mass component is first peak neutron-capture elements ($M_A$), and no neutron-capture elements are made in the $^{56}$Ni component.
    }
    \label{tab:modelXLa}
\end{table}

\end{document}